\documentclass[a4paper,11pt,twoside]{article}
\usepackage{amsthm}
\usepackage[T1]{fontenc}
\usepackage[utf8]{inputenc}
\usepackage[english]{babel}
\usepackage{amsmath}
\usepackage{amsfonts}
\usepackage{enumerate}
\usepackage{amssymb}
\usepackage[usenames]{color}
\usepackage{xcolor}
\usepackage{graphicx}
\usepackage{amsmath,dsfont}
\usepackage{fancyhdr}
\usepackage{hyperref}
\colorlet{darkblue}{blue!50!black}
\hypersetup{
    colorlinks,%
    citecolor=darkblue,%
    filecolor=red,%
    linkcolor=darkblue,%
    urlcolor=darkblue,%
    pdfnewwindow=true,%
    pdfstartview={FitH}
}
\usepackage{fourier}
\usepackage{amsfonts}
\usepackage{subfigure}
\usepackage{bm}
\usepackage[a4paper]{geometry}
\def\dz{dz}

\newtheorem{theo}{Theorem}[section]
\newtheorem{lemm}[theo]{Lemma}

\newtheorem{prop}[theo]{Proposition}

\newcommand{\R}{\mathbb{R}}
\numberwithin{equation}{section}

\newcommand{\C}{\mathbb{C}}

\newcommand{\ep}{\mathrm{ep}}

\def\X{{\bf X}}

\def\u{\textit {\textbf u}}
\def\<{\langle}
\def\>{\rangle}

\def\0{{\bf 0}}

\def\pup1{{\partial\u\over\partial x_1}}

\newcommand{\cD}{\mathcal{D}}
\newcommand{\cF}{\mathcal{F}}
\def\beq{\begin{equation}}
\def\eeq{\end{equation}}

\renewcommand{\Re}{\mathop{\mathrm{Re}}\nolimits}

\newcommand{\Ker}{\mathop{\mathrm{Ker}}\nolimits}
\newcommand{\Ran}{\mathop{\mathrm{Ran}}\nolimits}

\renewcommand{\sp}{\mathop{\mathrm{sp}}\nolimits}

\newcommand{\tr}{\mathop{\mathrm{tr}}\nolimits}

\newcommand{\cI}{\mathcal{I}}
\newcommand{\pI}{{\partial\mathcal{I}}}
\def\d{\mathrm{d}}
\def\e{\mathrm{e}}
\def\i{\mathrm{i}}
\newcommand{\PP}{\mathbb{P}}
\newcommand{\EE}{\mathbb{E}}

\newcommand{\cS}{\mathcal{S}}
\newcommand{\cE}{\mathcal{E}}
\newcommand{\cL}{\mathcal{L}}

\newcommand{\cQ}{\mathcal{Q}}

\newcommand{\txi}{{\widetilde{\xi}}}
\newcommand{\teta}{{\widetilde{\eta}}}

\title{A Detailed Fluctuation Theorem for Heat Fluxes\\ in Harmonic Networks out of Thermal Equilibrium}
\author{Mondher Damak$^1$ \and Mayssa Hammami$^{1,2}$ \and Claude-Alain Pillet$^2$}
\begin{document}
\def\today{}
\maketitle
\begin{center}
$^1$Departement of Mathematics, Faculty of Sciences of Sfax, University of Sfax, Tunisia.\\
$^2$Aix Marseille Univ, Université de Toulon, CNRS, CPT, Marseille, France
\end{center}

\vskip 1cm
{\small
{\bf Abstract.}
We continue the investigation, started in~\cite{JPS}, of a network of harmonic
oscillators driven out of thermal equilibrium by heat reservoirs. We study the
statistics of the fluctuations of the heat fluxes flowing between the network
and the reservoirs in the nonequilibrium steady state and in the large time
limit. We prove a large deviation principle for these fluctuations and derive
the fluctuation relation satisfied by the associated rate function.
\thispagestyle{empty}

 }

\section{Introduction}

{\sl Fluctuation Relations} (FRs for short) describe universal features of the
statistical properties of physical systems. The first instance of such a
relation goes back to 1905 and the celebrated work of Einstein on Brownian
motion. Despite a few early occurrences in the literature\footnote{
see~\cite{BK}, and also, in view of the now well understood connection with the
thermodynamic formalism, \cite[Proposition~5.3.2, Equ.~(3.9)]{Ru1}.}\!, it is
only after the works~\cite{ECM,ES,GC1,GC2} that FRs became a major research
direction in nonequilibrium statistical mechanics, see, e.g.,
\cite{Ja,Cr,LS,Ma,CG} on the theoretical side and~\cite{CCZ,CDF} on the
experimental one. See also~\cite{Ga2,Ga3,Ru2} for more mathematically oriented introductions
to the subject and  the reviews~\cite{RM,Se,JPR,JOPP} for more exhaustive references. 
In this work, we shall adhere to the somewhat restrictive but mathematically precise perspective
advocated by Gallavotti and Cohen in~\cite{GC1,GC2}, and call FR a universal
\,--\, i.e., model independent \,--\,  symmetry property of the rate function describing the
large deviations of some distinguished observable of a physical system in a
steady state in the large time limit. As explained in~\cite{ECM,Ga1,LS,JPR}, such
FRs, sometimes coined {\sl detailed FRs} in the physics literature, provide
extensions of the well-known Green--Kubo and Onsager relations of linear response
theory to the nonequilibrium regime.

While FRs in smooth chaotic dynamical systems on compact phase space are now
pretty well understood, their status is more problematic in the presence of
singularities and for general deterministic or stochastic dynamical system with
non-compact phase space.  As observed through the study of specific models, FRs
may only have a limited domain of validity and/or acquire non-universal features
in such circumstances, see~\cite{BGGZ,Fa,Vi,RT,RH}.

\begin{itemize}
\item From the physical point of view, there are issues related to the proper
choice of the relevant observable, as discussed in~\cite{JPS}. While most
studies concern entropy production, work and heat transfer in time-dependent
protocols involving a single heat reservoir, there is also some interests in
investigating individual heat fluxes in multi-reservoir systems.

\item From the mathematical point of view, the problems are often related to the
failure of standard approaches to the derivation of a large deviation principle
(see the series~\cite{JNPS1}--\cite{JNPS4} and~\cite{Ne}), or due to the
technical difficulties met in  applying the contraction principle, as
in~\cite[Section~3.4]{BL}.
\end{itemize}

In order to reach a better understanding of FRs we need to further investigate
simple models which allow for a clean mathematical treatment. Networks of
oscillators~\cite{MN,EZ} are among the simplest candidates. While some
progresses have been achieved in our understanding of the non-equilibrium
dynamics of  networks of anharmonic oscillators (see, e.g., the recent
works~\cite{CE,CEHR}), a complete picture of FRs for these systems seems to be
still out of reach of currently available techniques (to the best of our
knowledge, the only partial results can be found in~\cite{RT} for chains of
oscillators). The circumstances are much more favorable to networks of harmonic
oscillators~\cite{CDF,KSD}. Despite being very special, the latter provide
effective models for a wide range of systems and processes, from macroscopic
electrical circuits~\cite{CCZ,GMR} to the microscopic dynamics of
protein~\cite{HE,ADJ}, including the motion of mesoscopic colloidal
particles~\cite{Vi,JPC}. A novel control-theoretic approach to stochastically
driven harmonic networks has been developed in~\cite{JPS}. There, FRs were
obtained for various quantities related to the entropy produced by a general
harmonic network driven out of equilibrium by thermal forcing (see
Eq.~\eqref{Eq:EPFR} below for a typical result). The purpose of the present work
is to continue these investigations, following the same control-theoretic
strategy, and focusing on the individual energy currents flowing between the
network and its environment.

Let us briefly describe the settings of~\cite{JPS} which will be used in this
work. We focus on a collection, indexed by a finite set $\cI$, of
one-dimensional harmonic oscillators with position and momentum coordinates
$q=(q_i)_{i\in\cI}$  and $p=(p_i)_{i\in\cI}$. The Hamiltonian of this system is
the quadratic form
\[
H(q,p)=\frac12|p|^2+\frac12|\kappa q|^2,
\]
where $|\,\cdot\,|$ denotes the Euclidean norm  and $\kappa$ an automorphism of
$\R^\cI$.

Besides the conservative harmonic forces deriving from this Hamiltonian, a
subset of the oscillators, indexed by $\pI\subset\cI$, is acted upon by thermal
reservoirs. The latter are described by Langevin forces
\begin{equation}
f_i(p,q)=(2\gamma_i\vartheta_i)^{\frac{1}{2}}\dot{w}_i-\gamma_i p_i,\qquad(i\in\pI)
\label{Langevinforce}
\end{equation}
where $\vartheta_i>0$ denotes the temperature of the $i^\text{th}$--reservoir,
$\gamma_i>0$ the rate at which energy is dissipated in this reservoir and
$\dot{w}_i$ is a standard white noise. We interpret the work\footnote{The
integral there is to be taken in Itô's sense.}
\begin{equation}
\Phi_i(t)=\int_0^t\left[(2\gamma_i\vartheta_i)^\frac{1}{2} 
p_i(s)\dot{w}_i(s)+\gamma_i(\vartheta_i-p_i(s)^2)\right]\d s
\label{PhiiDef}
\end{equation}
performed by the Langevin force $f_i$ during the time interval $[0,t]$ as the
amount of heat injected in the network by the $i^\text{th}$--reservoir during this
period. We denote by $\Xi=\R^\pI$ the vector space where the heat currents
$\Phi(t)=(\Phi_i(t))_{i\in\pI}$ take their values and write the associated
Euclidean inner product as $\langle\xi,\Phi\rangle=\sum_{i\in\pI}\xi_i\Phi_i$.

The main results of the present work concern the statistics of the $\Xi$--valued
process $\{\Phi(t)\}_{t\ge0}$  induced by the {\sl stationary} Markov process
generated by the system of stochastic differential equations
\[
\dot q=\nabla_pH(q,p),\qquad\dot p=-\nabla_qH(q,p)+f,
\]
with appropriate initial conditions. More precisely, and under a controllability
condition which ensures the existence and uniqueness of an invariant measure for
this Markov process:

\begin{itemize}
\item We identify a subspace $\cL\subset\Xi$ characterized by the fact that for
$\xi\in\cL$ one has
\begin{equation}
\langle\xi,\Phi(t)\rangle=\cQ_\xi(q(t),p(t))-\cQ_\xi(q(0),p(0))
\label{xiW}
\end{equation}
where $\cQ_\xi$ is a quadratic form which is a first integral of the harmonic
network. Applying a general result of~\cite{BJP} allows us to describe the
asymptotics of~\eqref{xiW} in the limit $t\to\infty$.  Besides a large deviation
principle (LDP for short) for the fluctuations of order $t$ of this quantity, we
also get the explicit form of its limiting distribution, which has full support.
This is in sharp contrast with what would happen if $\cQ_\xi$ was a bounded
function on the phase space: the right-hand side of~\eqref{xiW}\,--\, often
coined a ``boundary term'' in the physics literature \,--\, would have no order $t$
fluctuations and its limiting law would have compact support.

\item We then focus on the component $\Phi(t)^\perp$ of the heat flux
orthogonal to $\cL$. We show that it has fluctuations of order $t$ satisfying a
local LDP whose rate function $I$ is the (partial) Legendre transform of a real
analytic function $g$ for which we provide several explicit representations. In
particular, we connect $g$ to the spectral properties of a finite dimensional
matrix and the domain of validity of the LDP to some associated algebraic
Riccati equation. Both functions $g$ and $I$ satisfy a FR.

\item We derive a simple sufficient condition, in terms of the solutions to the
above mentioned Riccati equation, which ensures that our LDP and the associated
FR hold globally. We also provide several examples where our condition is
fulfilled. This shows that there is a regime where the components of the heat
flux along the subspace $\cL$ are responsible for the failure of the global FR
for the entropy production observed in~\cite{JPS}. Our examples show, however,
that our sufficient condition does not survive strong thermal forcing.

\item In cases where the LDP for $\Phi(t)^\perp$ holds globally without our
sufficient condition being satisfied, we show that the rate function only
satisfies the universal FR on a proper subset of $\Xi$ which is again described
in terms of the solutions to the Riccati equation.
\end{itemize}

The remaining parts of the paper are organized as follows. In Section~\ref{SCT Model}, 
we introduce a general class of harmonic networks driven out of thermal
equilibrium by heat reservoirs. We describe the stochastic processes generated
by their nonequilibrium dynamics and, within this probabilistic framework, we
identify the fluxes of energy flowing between the network and the heat
reservoirs. Then, we briefly recall some results of~\cite{JPS} on the
fluctuations of the entropy produced by the networks in a nonequilibrium steady
state: a large deviation principle and the associated FRs. Finally, we sketch a
naive argument which will motivate the approach followed in this work.

In Section~\ref{SCT Results}, we formulate our main results on the fluctuations
of the heat fluxes in a nonequilibrium steady state of the network. Under a
natural controllability assumption, we provide an explicit formula for the large
time asymptotics of the cumulant generating function of these fluxes. Then, we
describe the resulting {\sl local} LDP for the fluctuations of the heat fluxes
and the associated FRs. Finally, under an additional assumption on the network,
we provide a {\sl global} LDP with its associated FRs.

In Section~\ref{SCT Examples}, we provide some specific examples to which our results apply.
The final Section~\ref{SCT Proofs} collects all the proofs of our results.

\medskip\noindent
{\bf Acknowledgements.} This research was supported by the Agence Nationale de
la Recherche (ANR) through the grant {\sl NONSTOPS} (ANR-17-CE40-0006) and the
CNRS collaboration grant {\sl Fluctuation theorems in stochastic systems}. The
work of C.-A.P. has been carried out in the framework of the Labex {\sl Archimède}
(ANR-11-LABX-0033) and of the {\sl A*MIDEX} project (ANR-11-IDEX-0001-02), funded by
the “Investissements d’Avenir” French Government program managed by the ANR.

Parts of this work were performed during the visits of M.D. and M.H. at the
University of Toulon and of C.-A.P. at the University of Sfax. We thank the CPT,
the University of Toulon and the Mathematics Department of Sfax for their
hospitality and support. M.H. and C.-A.P. are also grateful to the Centre de
Recherches Mathématiques de l'Université de Montréal for its hospitality and the
Simons foundation and CNRS for their support during their stay in Montréal in
the fall 2018.

\section{The model}
\label{SCT  Model}

\subsection{Setup}

In order to set up the notation to be used in the sequel, we briefly recall the
general framework of~\cite{JPS}, referring the reader to this paper for more
details.

\begin{quote} \textbf{Notations and conventions.} Let $E$ and $F$ be real or
complex Hilbert spaces. $L(E,F)$ denotes the set of (continuous) linear
operators $A:E\to F$ and $L(E)=L(E,E)$. For $A\in L(E,F)$, $A^\ast\in L(F,E)$
denotes the adjoint of $A$, $\|A\|$ its operator norm, $\Ran A\subset F$ its
range and $\Ker A\subset E$ its kernel. We denote the spectrum of $A\in L(E)$ by
$\sp(A)$. $A$ is non-negative (resp.\;positive), written $A \geq 0$ (resp.\;$A>
0$), if it is self-adjoint and $\sp(A)\in [0,\infty[$ (resp.
$\sp(A)\subset]0,\infty[$). We write $A\geq B$ whenever $A-B\geq0$. A pair
$(A,Q)\in L(E)\times L(F,E)$ is said to be controllable if the smallest
$A$-invariant subspace of $E$ containing $\Ran Q$ is $E$ itself. Denoting by
$\C_{\mp}$ the open left/right half-plane, $A\in L(E)$ is said to be
stable/anti-stable whenever $\sp(A)\subset\C_{\mp}$. 
\end{quote}

We consider the harmonic network described in the Introduction. The
configuration space $\R^\cI$ is endowed with its Euclidean structure and the
phase space $\Gamma=\R^\cI\oplus\R^\cI$ is equipped with its canonical
symplectic structure. On these spaces, $|\ \ |$ and $\,\cdot\,$ denote the
Euclidean norm and inner product, respectively. Recall that the Euclidean inner
product of the space $\Xi=\R^\pI$ is written
$\langle\xi,\Phi\rangle=\sum_{i\in\pI}\xi_i\Phi_i$.

\begin{quote}
\textbf{Convention.}  We identify $\xi\in\Xi$ with the element of $L(\Xi)$ defined by
\[
\xi:(u_i)_{i\in\pI}\mapsto(\xi_i u_i)_{i\in\pI}.
\]
In particular, whenever we write inequalities involving such $\xi$, they are always to be
interpreted as operator inequalities.
\end{quote}

Introducing the linear map $\iota\in L(\R^\pI,\R^\cI)$ defined by
\[
\iota:(u_i)_{i\in\pI}\mapsto(\sqrt{2\gamma_i}u_i)_{i\in\pI}\oplus0_{\R^{\cI\setminus\pI}},
\]
we set
\begin{equation}
x=\begin{bmatrix}p\\\kappa q\end{bmatrix},
\qquad
A=\begin{bmatrix}
-\frac12\iota\iota^\ast&-\kappa^\ast\\
\kappa&0
\end{bmatrix},
\qquad
Q=\begin{bmatrix}\iota\\0\end{bmatrix}\vartheta^{\frac12}.
\end{equation}
The internal energy of the network then writes $h(x)=\frac12|x|^2$, and its
dynamics is described by the following system of Itô stochastic differential
equations
\begin{equation}
\d x(t)= A x(t) \d t + Q \d w(t),
\label{eqmotionx}
\end{equation}
$w$ denoting a standard $\Xi$--valued Wiener process. The solution of the Cauchy
problem associated to~\eqref{eqmotionx}, with initial condition $x(0)=x_0$, can
be written explicitly as
\begin{equation}
x(t)=\e^{tA}x_0+\int_0^t\e^{(t-s)A}Q\d w(s).
\label{solteqmotion}
\end{equation}
This relation defines a family of $\Gamma$--valued Markov processes indexed by
the initial condition $x_0\in\Gamma$. The generator of this degenerate diffusion
process is given by
\begin{equation}
L=\frac12\nabla\cdot B\nabla+Ax\cdot\nabla,
\label{LDef}
\end{equation}
where $B=QQ^\ast$. We shall denote $\PP_{x_0}$ the probability measure induced
on the path space $C([0,\infty[,\Gamma)$ and by $\EE_{x_0}$ the corresponding
expectation functional. Given a probability measure $\nu$ on $\Gamma$, we
further set $\PP_\nu=\int\PP_x\nu(\d x)$ and define similarly $\EE_\nu$.

For later references, we note the following structural relations
\begin{equation}
\Ker(A-A^\ast)=\{0\}, \quad A+A^\ast=-Q\vartheta^{-1}Q^\ast,
\quad Q^\ast Q>0,\quad [\vartheta,Q^\ast Q]=0.
\label{structuralconstraints1}
\end{equation}
Moreover, denoting by $\theta$ the time-reversal involution $(p,q)\mapsto(-p,q)$ of $\Gamma$,
\begin{equation}
\theta=\theta^\ast=\theta^{-1},\quad\theta Q=-Q,\quad\theta A\theta=A^\ast.
\label{structuralconstraints2}
\end{equation}
Setting
\[
\Omega=\frac12(A-A^\ast),
\]
we also have
\begin{equation}
\theta B\theta=B^\ast=B,\quad
\theta\Omega\theta=\Omega^\ast=-\Omega.
\label{structuralconstraints3}
\end{equation}
In the sequel, we shall further assume the Kalman condition
\begin{quote}
{\bf (C)} The pair $(A, Q)$ is controllable.
\end{quote}
We recall (see \cite[Theorem 3.2]{JPS} and references therein) that under this
assumption the process~\eqref{solteqmotion} admits a unique invariant measure
$\mu$, the centered Gaussian measure on $\Gamma$ with covariance
\begin{equation}
M=\int_0^{\infty}\e^{sA}B\e^{sA^\ast}\d s,
\label{Mdef}
\end{equation}
which is also characterized as the unique solution of the Lyapunov equation 
$AM+MA^\ast+B=0$. 

\medskip\noindent
{\bf Remark.} If the environment is in thermal equilibrium at temperature $T_0>0$, 
i.e., if $\vartheta=T_0 I_{\Xi}$, then $M=T_0I_\Gamma$ and $\mu$ is the Gibbs 
measure $\mu(\d x)\propto\e^{-h(x)/T_0}\d x$. 

\subsection{Heat fluxes and entropy production}

Following~\cite{JPS}, we shall interpret the work~\eqref{PhiiDef}  performed by
the Langevin force during the time interval $[0,t]$ as the amount of heat
injected in the network by the $i^\text{th}$--reservoir during this period.
In~\cite{JPS}, a LDP and extended fluctuation relations were proven for the
total amount of entropy dissipated in the reservoirs, i.e., the entropy produced
by the network
\beq
\mathfrak{S}(t)=-\langle\vartheta^{-1},\Phi(t)\rangle.
\label{SDef}
\eeq
Let us briefly recall these results.

Assuming the Kalman condition {\bf (C)}, the family $\{\mathfrak{S}(t)\}_{t\ge0}$ satisfies
a global LDP with a good rate function
$I:\R\to[0,+\infty]$, i.e., for any Borel set $S\subset\R$, one has
\[
-\inf_{s\in\dot{S}}I(s)\le
\liminf_{t\to\infty}\frac1t\log\PP_\mu[t^{-1}\mathfrak{S}(t)\in S]\le
\limsup_{t\to\infty}\frac1t\log\PP_\mu[t^{-1}\mathfrak{S}(t)\in S]\le
-\inf_{s\in\bar S}I(s)
\]
where $\dot S$ and $\bar S$ denote respectively the interior and the closure of $S$.
The mean entropy production rate
\begin{equation}
\ep=\lim_{t\to\infty}\frac1t\EE_\mu[\mathfrak{S}(t)]
\label{epDEF}
\end{equation}
exists and is non-negative. Whenever $\ep>0$, the rate function satisfies the FR
\beq
I(-s)-I(s)=s
\label{Eq:EPFR}
\eeq
for $|s|\le\ep$. However, this universal relation fails for $|s|>\ep$ where, instead, a model 
dependent {\sl extended fluctuation relation} holds (see~\cite[Section~3.7]{JPS}).

Our aim here is to derive a LDP and FRs for the individual heat fluxes
$\Phi(t)=(\Phi_i(t))_{i\in\pI}$. To motivate our approach, let us sketch a naive
argument leading to the desired results.

The Gärtner-Ellis theorem (see, e.g., \cite{DZ} or~\cite{dH}) is a well-known
route to the LDP. To follow it, one has to show the existence and some
regularity properties of the large time limit
\begin{equation}
e(\xi)=\lim_{t\to\infty}\frac1t g_t(\xi)
\label{eDef}
\end{equation}
of the cumulant generating function
\begin{equation}
g_t(\xi)=\log\EE_\mu\left[\e^{\langle\xi,\Phi(t)\rangle}\right].
\label{Eq:gtxi}
\end{equation}
A simple calculation yields the expression
\begin{equation}
\langle\xi,\Phi(t)\rangle=
\frac{t}{2}\tr(Q\xi Q^\ast)+\int_0^t\left[x(s)\cdot Q\xi\d w(s)
-\frac12 x(s)\cdot Q\vartheta^{-1/2}\xi\vartheta^{-1/2} Q^\ast x(s)\d s
\right].
\label{funcTDEx}
\end{equation}
By Itô calculus, one has
\[
\d(\e^{\langle\xi,\Phi(t)\rangle}f(x(t)))=\e^{\langle\xi,\Phi(t)\rangle}
\left[(L_\xi f)(x(t))\d t
+\langle\xi Q^\ast x(t) f(x(t))+Q^\ast\nabla f(x(t)),\d w(t)\rangle\right],
\]
where
\[
L_{\xi}=\frac12\nabla\cdot B\nabla+A_{\xi}x\cdot\nabla-\frac12 x\cdot C_\xi x
+\frac12\tr(Q\xi Q^\ast),
\]
is a deformation of the Markov generator~\eqref{LDef}, the matrices $A_\xi$ and
$C_\xi$ being given by
\begin{equation}
A_\xi=A+Q\xi Q^\ast,\qquad
C_\xi=Q\xi(\vartheta^{-1}-\xi)Q^\ast.
\label{AxiDef}
\end{equation}
A naive application of Girsanov formula yields
\[
\EE_x\left[\e^{\langle\xi,\Phi(t)\rangle}f(x(t))\right]=\left(\e^{tL_\xi}f\right)(x),
\]
and in particular
\[
g_t(\xi)=\log\int(\e^{tL_\xi}1)(x)\mu(\d x).
\]
Given the specific form of $L_\xi$ and the fact that it generates a positivity
preserving semigroup, it is natural to seek an eigenvector $\Psi_\xi$ to its
dominant eigenvalue
$\lambda_{\xi}=\max\{\Re\lambda\,|\,\lambda\in\sp(L_{\xi})\}$ in the Gaussian
form
\[
\Psi_\xi(x)=\e^{-\frac12 x\cdot X_{\xi}x}.
\]
A simple calculation shows that the eigenvalue problem splits into the following
algebraic Riccati equation for a symmetric matrix $X$,
\beq
\mathcal{R}_\xi(X)\equiv X BX-X A_\xi-A^\ast_\xi X-C_\xi=0,
\label{Eq:RiccatiFirst}
\eeq
and the relation
\[
\lambda_\xi=\frac12\tr(Q\xi Q^\ast-BX_\xi)
\]
where $X_\xi$ denotes the {\sl maximal\/}\footnote{$X_\xi$ is maximal whenever
$X\le X_\xi$ for all self-adjoint $X$ such that $\mathcal{R}_\xi(X)=0$.}
solution of~\eqref{Eq:RiccatiFirst}. From the structural
relations~\eqref{structuralconstraints1}--\eqref{structuralconstraints3} and the
fact that $[\xi,\vartheta]=0$ one easily deduces that the formal adjoint
$L^\ast_\xi$ of $L_\xi$ is given by
\begin{equation}
L^\ast_\xi=\Theta L_{\vartheta^{-1}-\xi}\Theta,
\label{Lxiadjointrelation}
\end{equation}
where the map $\Theta$ is defined by $\Theta f=f\circ\theta$.

Assuming $L_\xi$ to have a non-vanishing spectral gap, we obtain
\[
\int(\e^{tL_\xi}1)(x)\mu(\d x)=\e^{t\lambda_\xi}\left(d_\xi+o(1)\right)
\]
as $t\to\infty $, and hence
\[
\lim_{t\to\infty}\frac1t g_t(\xi)=\lambda_\xi,
\]
provided the prefactor
\beq
d_\xi=\frac{\int\Psi_{\vartheta^{-1}}(\theta x)\Psi_{\vartheta^{-1}-\xi}(\theta y)\Psi_\xi(x)
\d x\d y}
{\int\Psi_{\vartheta^{-1}}(\theta x)\Psi_{\vartheta^{-1}-\xi}(\theta y)\Psi_\xi(y)\d x\d y}
=\frac{\det(X_{\vartheta^{-1}})^{1/2}\det(X_\xi+\theta X_{\vartheta^{-1}-\xi}\theta)^{1/2}}
{\det(X_\xi+\theta X_{\vartheta^{-1}}\theta)^{1/2}\det(X_{\vartheta^{-1}-\xi})^{1/2}}
\label{dxiDef}
\eeq
is finite and positive.\footnote{Here, we used the fact that the steady state covariance $M$
satisfies $M^{-1}=\theta X_{\vartheta^{-1}}\theta$.}  The fluctuation relation
\beq
\lambda_{\vartheta^{-1}-\xi}=\lambda_\xi
\label{gFR}
\eeq
then follows from~\eqref{Lxiadjointrelation}.

Assuming the limiting cumulant generating function $\xi\mapsto\lambda_\xi$ to be everywhere
differentiable on $\Xi$, the Gärtner-Ellis theorem yields the LDP
\[
-\inf_{\varphi\in\dot{F}}I(\varphi)\le
\liminf_{t\to\infty}\frac1t\log\PP_\mu[t^{-1}\Phi(t)\in F]\le
\limsup_{t\to\infty}\frac1t\log\PP_\mu[t^{-1}\Phi(t)\in F]\le
-\inf_{\varphi\in\bar F}I(\varphi),
\]
where $\dot{F}/\bar F$ denotes the interior/closure of the Borel set $F\subset\Xi$,
the rate function being given by the Legendre transform
\[
I(\varphi)=\sup_{\xi\in\Xi}(\langle\xi,\varphi\rangle-\lambda_\xi).
\]
Relation~\eqref{gFR} thus translates into the FR
\begin{equation}
I(-\varphi)-I(\varphi)=-\langle\vartheta^{-1},\varphi\rangle,
\label{IFR}
\end{equation}
where we recognize, in the right-hand side, the entropy production rate
corresponding to the heat flux $\varphi$.

There are several issues with the above formal derivation. In particular, we
can't expect the fluctuation relation~\eqref{IFR} (resp.~\eqref{gFR}) to hold
for all values of $\varphi\in\Xi$ (resp.\;for all values of $\xi\in\Xi$).
Indeed, by the contraction principle, the validity of~\eqref{IFR} for all
$\varphi\in\Xi$ would entail the validity of~\eqref{Eq:EPFR} for all $s\in\R$,
in contradiction with the above mentioned result of~\cite{JPS}. The main
contribution of the present work is to provide a rigorous proof of a large
deviation principle for heat fluxes, an explicit formula for the rate function
$I$ and a description of the domain of validity of the universal
relation~\eqref{IFR}.

\section{Main results}
\label{SCT Results}

\subsection{The limiting cumulant generating function}

In this paragraph, we first formulate the generalized detailed balance relation
(see~\cite[Section~4.3]{EPR} and~\cite[Section~2.2]{BL}) which plays a central
role in our analysis. Then we state our main result on the large time
asymptotics of the cumulant generating function~\eqref{Eq:gtxi}.

\begin{prop}\label{propxitild}
Given $\xi\in\Xi$, we shall write $\txi\rhd\xi$ whenever $\txi\in L(\Gamma)$ is
self-adjoint and satisfies
\begin{equation}
\txi Q=Q\xi,\quad\theta\txi\theta=\txi.
\label{eqtildxi}
\end{equation}
To such a $\txi$, we associate the quadratic forms
\[
\cQ_\txi(x)=\frac12x\cdot\txi x,
\]
and
\[
\sigma_{\txi}(x)=\frac12 x\cdot\Sigma_{\txi}x,\qquad
\Sigma_{\txi}=[\Omega,\txi],
\]
and the measure $\mu_{\txi}$ on $\Gamma$ defined by
\[
\frac{\d\mu_\txi}{\d x}(x)=\e^{-\cQ_\txi(x)}.
\]
Then, the following assertions hold:
\begin{enumerate}[(1)]
\item $\mu_\txi$ and $\sigma_\txi$ satisfy
\[
\mu_\txi\circ\theta=\mu_\txi,\qquad
\sigma_\txi\circ\theta=-\sigma_\txi.
\]
\item Denote by $L^\txi$  the formal adjoint of the Markov
generator~\eqref{LDef} w.r.t.\;the inner product of the Hilbert space
$L^2(\Gamma,\mu_\txi)$. Then the generalized detailed balance relation
\[
\Theta L^\txi\Theta=L+\sigma_\txi
\]
holds.
\item There exists $\txi\in L(\Gamma)$ satisfying~\eqref{eqtildxi} and such that 
$\Sigma_\txi=0$ iff
\[
\e^{-\cQ_\txi(x)}L_\eta\e^{\cQ_\txi(x)}=L_{\eta+\xi}
\]
holds for all $\eta\in\Xi$. Moreover, under Condition~\textbf{(C)}, such a $\txi$,
if it exists, is unique and satisfies
\[
\sp(\txi)=\sp(\xi).
\]
\item The functional~\eqref{funcTDEx} can be written as
\[
\langle\xi,\Phi(t)\rangle=\cQ_\txi(x(t))-\cQ_\txi(x(0))+\int_0^t\sigma_\txi(x(s))\d s.
\]
\end{enumerate}
\end{prop}

Given the structure of the map $Q$ and the diagonal nature of $\xi$, the
existence of $\txi\in L(\Gamma)$ satisfying~\eqref{eqtildxi} is obvious. Apart
from Part~(3), the proof of the previous proposition is identical to the
elementary proof of Proposition~3.5 in~\cite{JPS} and we omit it. The first
statement in~(3)  follows from an explicit calculation. One easily checks that
$\Sigma_\txi=0$ is equivalent to $[A,\txi]=0$, which implies that $\txi
A^nQ=A^nQ\xi$ for any $n\ge0$. Thus, Condition~\textbf{(C)} immediately yields
the uniqueness of $\txi$.  Similarly, one deduces from the relation
$(\txi-z)^{-1}A^nQ=QA^n(\xi-z)^{-1}$, obviously valid for $z\in\C\setminus\R$,
that $\txi$ and $\xi$ have the same spectrum.

\begin{prop}\label{propgxi}
Assume that Condition {\bf (C)} holds.
\begin{enumerate}[(1)]
\item For $\xi\in\Xi$ and $\omega\in\R$, the operator
\begin{equation}
E_\xi(\omega)=Q^\ast(A^\ast-\i\omega)^{-1}\Sigma_\txi(A+i\omega)^{-1}Q
\label{Eomega}
\end{equation}
is self-adjoint on the complexification of\, $\Xi$ and does not depend on the
choice of $\,\txi\rhd\xi$. Moreover, the map $\R\times\Xi\ni(\omega,\xi)\mapsto
E_\xi(\omega)$  is continuous.
\item The set
\[
\cD=\bigcap_{\omega\in\R}\{\xi\in\Xi\,|\,I-E_{\xi}(\omega)>0\}
\]
is open, convex, centrally symmetric around the point $(2\vartheta)^{-1}$ and contains
\[
\cD_0=\{\xi\in\Xi\,|\,0<\xi<\vartheta^{-1}\}.
\]
Its lineality space\footnote{The lineality space of a convex set $\mathcal{C}\subset\R^n$
is the set of vectors $y\in\R^n$ such that $x+\lambda y\in\mathcal{C}$ for all
$x\in\mathcal{C}$ and all $\lambda\in\R$, see~\cite{Ro}} is given by
\beq
\cL=\bigcap_{\omega\in\R}\{\xi\in\Xi\,|\,E_\xi(\omega)=0\}
=\{\xi\in\Xi\,|\,\Sigma_\txi=0\text{ for some } \txi\rhd\xi\},
\label{DlinForm}
\eeq
and in particular $\boldsymbol{1}=(1,1,\ldots,1)\in\cL$.
\item The function
\[
g(\xi)=-\int_{-\infty}^{+\infty}\log\det(I-E_\xi(\omega))\frac{\d\omega}{4\pi},
\]
is convex and real analytic on $\cD$. It is centrally symmetric w.r.t.\;the
point $(2\vartheta)^{-1}$ and translation invariant in the direction $\cL$, i.e.,
\begin{equation}\label{G-Csymmetry}
g(\vartheta^{-1} - \xi)=g(\xi)=g(\xi+\eta)
\end{equation}
for all $\xi\in\cD$ and $\eta\in\cL$. In particular, $g(0)=g(\vartheta^{-1})=0$.
\item One has
\[
\cL=\nabla g(\cD)^\perp,
\]
and the following alternative holds: Either $\cD=\cL=\Xi$ and $g$ vanishes
identically, or $\cL^\perp\not=\{0\}$ and $g$ is strictly convex on the section
$\cS=\cD\cap\cL^\perp$, the closure of $\cS$ being a compact convex subset of
$\cL^\perp$.
\item For $\xi\in\Xi$, define
\begin{equation}
K_{\xi}=\begin{bmatrix}
-A_\xi&QQ^\ast\\
C_\xi&A^\ast_\xi\\
\end{bmatrix},
\label{Hamiltonianmatrix}
\end{equation}
where $A_{\xi}$ and $C_{\xi}$ are given by~\eqref{AxiDef}. Then $\cD$ is
the connected component of the point $\xi=0$ in the set
\[
\{\xi\in\Xi\,|\,\sp(K_{\xi})\cap\i\R=\emptyset\}.
\]
Moreover, for any $(\omega,\xi)\in\R\times\cD$ one has
\[
\det(K_{\xi}-\i\omega)=|\det(A+\i\omega)|^2\det(I-E_{\xi}(\omega)).
\]
\item The function $g$ has a bounded continuous extension to the closed set
$\,\overline{\cD}$ which is given by
\begin{equation}\label{eqgalg}
g(\xi)=\frac14\tr(Q\vartheta^{-1}Q^*)
-\frac14\sum_{\lambda\in\sp(K_\xi)}|\Re\lambda|m_\lambda,
\end{equation}
where $m_\lambda$ denotes the algebraic multiplicity of $\lambda\in\sp(K_\xi)$.
\item For any finite $\xi_0\in\partial\cD$ one has
\[
\lim_{\cD\ni\xi\to\xi_0}|\nabla g(\xi)|=\infty.
\]
Thus, setting $g(\xi)=+\infty$ for $\xi\in\Xi\setminus\overline{\cD}$ yields an
essentially smooth, essentially strictly convex, closed, proper convex function
$g:\Xi\to]-\infty,+\infty]$.
\item For all $\xi\in\overline{\cD}$ the Riccati
equation~\eqref{Eq:RiccatiFirst} has a maximal self-adjoint solution $X_\xi$.
The map $\xi\mapsto X_\xi$ is continuous and concave on $\overline{\cD}$, and
\beq
g(\xi)=-\frac12\tr(Q^\ast(X_\xi-\txi) Q).
\label{gxiForm}
\eeq
Moreover, setting $D_\xi=A_\xi-BX_\xi$,  the pair $(D_\xi,Q)$  is controllable
and $\sp(D_\xi)=\sp(K_\xi)\cap\overline{\C_-}$.
\end{enumerate}
\end{prop}

The lineality subspace $\cL$ is related to conservation laws of the harmonic
network. Indeed, under its Hamiltonian dynamics, the network evolves according
to $x_t=\e^{t\Omega}x_0$ and hence
\[
\frac{\d\ }{\d t}\cQ_\txi(x_t)=\sigma_\txi(x_t)=0
\]
for $\xi\in\cL$. It follows that, for any $\xi\in\cL$, the quadratic  form
$\cQ_\txi$ is a first integral of the Hamiltonian flow. In particular the
direction $\boldsymbol{1}\in\cL$ and the invariance
$g(\xi+\lambda\boldsymbol{1})=g(\xi)$ is related to the conservation of the
total energy of the network, $h=\cQ_I$. This symmetry of the cumulant generating
function of currents was already described, in the quantum setting,
in~\cite{AGMT}, see also~\cite{BPP} for a detailed discussion.

It follows from~\cite[Theorem~3.13]{JPS} that the entropy production rate of the
network~\eqref{epDEF} is related to the function $g$ by
\[
\ep=-\langle\vartheta^{-1},\nabla g(0)\rangle.
\]
Thus, $\ep=0$ whenever the alternative $\cL=\Xi$ in Part~(4) of
Proposition~\ref{propgxi} holds. In the following we shall avoid trivialities
assuming, without further notice, that $\ep>0$ and hence $\cL^\perp\not=\{0\}$
and
\[
\cD=\cS\oplus\cL
\]
with $\cS=\cD\cap\cL^\perp$.
\medskip
By Proposition~\ref{propgxi}~(8) the functions
\[
\Lambda_-(\xi)=-\min\sp(X_\xi+\theta X_{\vartheta^{-1}}\theta),\qquad
\Lambda_+(\xi)=\min\sp(X_{\vartheta^{-1}-\xi}),
\]
are continuous and respectively convex/concave on $\overline{\cD}$. The function
$g$ of the preceding proposition is related to the limiting cumulant generating
function~\eqref{eDef} by the following
\begin{prop}\label{proplimgxi}
Under Assumption {\bf(C)} one has
\begin{equation}
e(\xi)=\lim_{t\to\infty}\frac1t g_t(\xi) =
\begin{cases}
g(\xi)&\text{for }\xi\in\cD_\infty\\[4pt]
+\infty&\text{for }\xi\in\Xi\setminus\overline{\cD_\infty},
\end{cases}
\label{limgxig}
\end{equation}
where (compare this with the right-hand side of~\eqref{dxiDef})
\[
\cD_\infty=\{\xi\in\cD\,|\,\Lambda_-(\xi)<0<\Lambda_+(\xi)\}
\]
is a bounded, open, convex subset of  $\,\cD$ such that
\[
\overline{\cD_0}\setminus\{\vartheta^{-1}\}\subset\cD_\infty.
\]
In particular, $\cD_\infty$ contains a neighborhood of\, $0$.
\end{prop}

\subsection{Fluctuations of conserved quantities}

As mentioned above, each $\xi\in\cL$ is associated to a first integral
$\cQ_\txi$ of the harmonic network. In this section, we briefly focus on these
conserved quantities. Since $\xi_\lambda=\xi+\lambda\boldsymbol{1}\in\cL$ and
$\txi_\lambda=\txi+\lambda I>0$ for $\lambda\in\R$ large enough, there is no
loss of generality in assuming that $\cQ_\txi\ge0$. It follows from
Proposition~\ref{propxitild}~(4) and~\cite[Proposition~2.2]{BJP} that the law of
\[
\langle\xi,\Phi(t)\rangle=\cQ_\txi(x(t))-\cQ_\txi(x(0))
\]
under $\PP_\mu$ converges towards a {\sl variance-gamma} distribution
\[
\lim_{t\to\infty}\PP_\mu[\langle\xi,\Phi(t)\rangle\in S]=\int_S f_\mathrm{vg}(q)\d q,
\]
with density
\[
f_\mathrm{vg}(q)
=|q|^{(m-1)/2}\int_{S^{m -1}} K_{(m-1)/2}\left(\frac{|q|}{|Nk|}\right)
\frac{\d\sigma(k)}{(2\pi|Nk|)^{(m+1)/2}},
\]
where\footnote{Recall that $M$, given in~\eqref{Mdef}, is the covariance of the
invariant measure $\mu$.}
$m=2|\cI|$, $N=\txi^{1/2}M\txi^{1/2}$, $\sigma$ is the Lebesgue measure on the
unit sphere $S^{m-1}$ of $\Gamma$, and $K$ denotes a modified Bessel function~\cite{W}.
As mentioned in the Introduction, the variance-gamma distribution as full support on $\R$.
Moreover, the above convergence is accompanied by a LDP: for any open
set $O\subset\R$,
\[
\lim_{t\to\infty}\frac1t\log\PP_\mu[t^{-1}\langle\xi,\Phi(t)\rangle\in O]=-\inf_{q\in O}I(q)
\]
with the rate function
\[
I(q)=\frac{|q|}{\max\sp(N)}.
\]
This applies, in particular, to the fluctuations of the total energy which was the primary
concern in~\cite{BJP}.

\subsection{A local Fluctuation Theorem}

While the results of the previous section quantify departures from the conservation laws,
the main results of this paper deal with the component of the heat currents fluctuations which 
do not violate these conservation laws.

Recall that $\cS=\cD\cap\cL^\perp$ is the (precompact, convex) base of the
cylinder $\cD\subset\Xi$. Denoting by $\Pi\in L(\Xi)$ the orthogonal projection
on $\cL^\perp$, setting $\cS_\infty=\Pi\cD_\infty$ and defining the function
$I:\cL^\perp\to[0,+\infty[$ by
\beq
I(\varphi)=\sup_{\xi\in\cS_\infty}\left(\langle\xi,\varphi\rangle-g(\xi)\right),
\label{IDef}
\eeq
a direct application of the Gärtner-Ellis theorem yields the following

\begin{theo}\label{theolocalLDP}
Assume that Condition {\bf (C)} holds. Then, under the law\,
$\PP_\mu$, the family $\{\Pi\Phi(t)\}_{t\ge0}$ satisfies a local
LDP with the good rate function $I$ given by~\eqref{IDef}, i.e., for any Borel set 
$F\subset\cL^\perp$, one has
\begin{equation}
-\inf_{\varphi\in\dot{F}\cap\cF}I(\varphi)\le
\liminf_{t\to\infty}\frac1t\log\PP_\mu[t^{-1}\Pi\Phi(t)\in F]\le
\limsup_{t\to\infty}\frac1t\log\PP_\mu[t^{-1}\Pi\Phi(t)\in F]\le
-\inf_{\varphi\in\bar F}I(\varphi),
\label{LDF}
\end{equation}
where $\dot F$ and $\bar F$ denote respectively the interior and the closure of $F$ and
\[
\cF=\nabla g(\cS_\infty).
\]
Moreover, for $\varphi\in\cF_0=\{\nabla g(\xi)\,|\,\xi\in\cS_\infty\text{ and }
\vartheta^{-1}-\xi\in\cS_\infty\}\supset\nabla g(\cD_0)$, the fluctuation relation
\[
I(-\varphi)=I(\varphi)-\langle\vartheta^{-1},\varphi\rangle,
\]
holds.
\end{theo}

\subsection{A global Fluctuation Theorem}

To improve on Theorem~\ref{theolocalLDP} and obtain a global LDP on $\cL^\perp$,
we impose a further condition on the network:
\begin{quote}
{\bf (R)}   
$
\displaystyle
\min_{\xi\in\partial\cS}\left(\Lambda_+(\xi)-\Lambda_-(\xi)\right)>0.
$
\end{quote}
Since $\Lambda_+-\Lambda_-$ is a concave function of $\xi$ on $\cS$,
Condition~\textbf{(R)} ensures that it is positive on $\cS$.
\begin{theo}\label{theoglobalLDP}
Assume that Conditions {\bf (C)} and {\bf (R)}  hold. Then, under the law
$\,\PP_\mu$, the family $\{\Pi W(t)\}_{t\ge0}$ satisfies a global
LDP with the good rate function $I$ given by~\eqref{IDef}, i.e., for any Borel set 
$F\subset\cL^\perp$, one has
\begin{equation}
-\inf_{\varphi\in\dot{F}}I(\varphi)\le
\liminf_{t\to\infty}\frac1t\log\PP_\mu[t^{-1}\Pi W(t)\in F]\le
\limsup_{t\to\infty}\frac1t\log\PP_\mu[t^{-1}\Pi W(t)\in F]\le
-\inf_{\varphi\in\bar F}I(\varphi),
\label{LDFagain}
\end{equation}
where $\dot F$ and $\bar F$ denote respectively the interior and the closure of $F$. Moreover, 
the fluctuation relation
\beq
I(-\varphi)=I(\varphi)-\langle\vartheta^{-1},\varphi\rangle,
\label{FRatlast}
\eeq
holds for all $\varphi\in\cL^\perp$.
\end{theo}

\noindent{\bf Remark.} We stress that Theorem~\ref{theoglobalLDP} only gives
sufficient conditions for the global validity of the LDP. We conjecture that
Condition~{\bf (C)}  alone is sufficient for~\eqref{LDFagain} to hold for all
Borel sets $F\subset\cL^\perp$ (with the rate function $I$ given by~\eqref{IDef}).
However, we were not able to prove this claim,
and in particular we are not aware of any general result in the theory of large
deviations which would imply it. We leave this conjecture as an interesting
open problem.

\medskip\noindent{\bf Discussion.} 
1. The quantity $-\langle\vartheta^{-1},\varphi\rangle$ is the entropy production rate
associated with a given heat current $\varphi\in\cL^\perp$ (compare with~\eqref{SDef}). 
Thus, the FR~\eqref{FRatlast} implies that current fluctuations $\varphi$ with negative entropy
production rate are exponentially suppressed, as $t\to\infty$, relative to the opposite
fluctuation $-\varphi$ (which has positive entropy production rate).

2. We note that, according to the results of~\cite{JPS}, the entropy
production~\eqref{SDef} {\sl never} satisfies the FR~\eqref{Eq:EPFR} globally
(i.e., for all $s\in\R$). This is in sharp contrast with the component of the
heat flux along $\cL^\perp$. Indeed, we will see in the next section that a
global FR is possible in this case.

\noindent 3. Suppose that, in view of the above conjecture, Condition~{\bf (R)}
being violated, the LDP~\eqref{LDFagain} holds for all Borel sets
$F\subset\cL^\perp$. It follows that the FR~\eqref{FRatlast} does not hold for all
$\varphi\in\cL^\perp$ but is replaced by an extended \,--\, i.e., non-universal
\,--\, FR on the unbounded set
\[
\{\varphi\in\cL^\perp\,|\,\text{either }\varphi\not\in\nabla g(\cS_\infty)
\text{ or }-\varphi\not\in\nabla g(\cS_\infty)\},
\]
as illustrated in Figure~\ref{RateCumulLozenge}, below. The graph of the rate
function in the region $\cL^\perp\setminus\nabla g(\cS_\infty)$ has the peculiar
form of a ruled surface. More precisely, for $\xi\in\partial\cS_\infty$ let
$\varphi_0=\nabla g(\xi)$ and denote by $\eta$ the exterior normal to
$\cS_\infty$ at $\xi$. Then, one has
$I(\varphi_0+\lambda\eta)=I(\varphi_0)+\lambda\eta\cdot\xi$ for all $\lambda>0$.
This behavior generalizes to the multivariate case the affine character of the
rate function of scalar observables (heat, entropy production,\dots) found
in~\cite{CZ,BJTM,Fa,RH,Vi,JPS}. In these circumstances, the rate function is
convex on $\cL^\perp$ but fails to be strictly convex on the complement of
$\nabla g(\cS_\infty)$. As in the scalar cases mentioned above, the somewhat
striking consequence of these {\sl extended FRs\/} is an increase of the probability ratio for
current fluctuations with negative to positive entropy production rate.

\noindent 4. From a mathematical perspective, deviations from the
FR~\eqref{FRatlast} are due to the lack of essential smoothness of the limiting
cumulant generating function~\eqref{limgxig}. In the setting of the present
work, this can be traced back to the divergence of a pre-exponential factor in
the asymptotic expansion of the cumulant generating function (the term $d_t$
in~\eqref{PreExp} below).  The analysis of~\cite{EN,RH} suggests that this
phenomenon may remain relevant beyond the harmonic/Gaussian setting, however we
are not aware of any rigorous result in this direction.

\noindent 5. From the physical point of view, the circumstances leading to the failure of the 
global FR~\eqref{FRatlast} are still not well understood. However, the examples below tend to 
indicate that the strength of the thermal drive (and hence large currents) is one determinant factor.


\section{Examples}
\label{SCT Examples}

Observe that Eq.~\eqref{Eomega} implies that the matrix $E_\xi(\omega)$ and
thence the function $g(\xi)$ are invariant under the simultaneous rescaling
\[
\vartheta_i\mapsto\lambda\vartheta_i,\qquad
\xi_i\mapsto\lambda^{-1}\xi_i
\]
with $\lambda>0$. Furthermore, one easily checks that, under the same rescaling,
the maximal solution to the Riccati equation~\eqref{Eq:RiccatiFirst} obeys
$X_\xi\mapsto\lambda X_\xi$, so that
$\Lambda_\pm(\xi)\mapsto\lambda\Lambda_\pm(\xi)$. Consequently, without losing
in generality, we shall fix the average temperature according to
\[
\frac1{\vert\pI\vert}\sum_{i\in\pI}\vartheta_i^{-1}=1
\]
 in all our examples, denoting temperature ratios by
 $[\vartheta_1:\vartheta_2:\cdots]$. For systems out of thermal equilibrium, we
 shall also use a special system of cartesian coordinates on the space
 $\cL^\perp$: we set its origin at the orthogonal projection of the symmetry
 center $\left((2\vartheta_i)^{-1}\right)_{i\in\pI}$ on $\cL^\perp$, and chose
 the first basis vector along the same direction. Finally, we note that in all
 the examples below, it is straightforward to check that Condition~\textbf{(C)}
 is satisfied and that $\cL=\R\boldsymbol{1}$. We shall therefore concentrate
 our discussions on the validity of Condition~\textbf{(R)}.

\subsection{A lozenge network}

\begin{figure}[h]
\begin{center}
\includegraphics[scale=0.75]{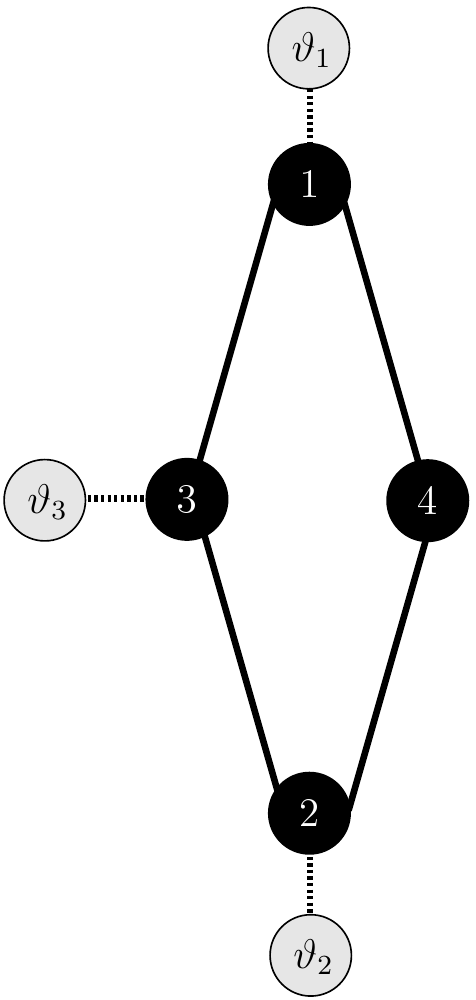}
\hskip 2.5cm
\includegraphics[scale=0.3]{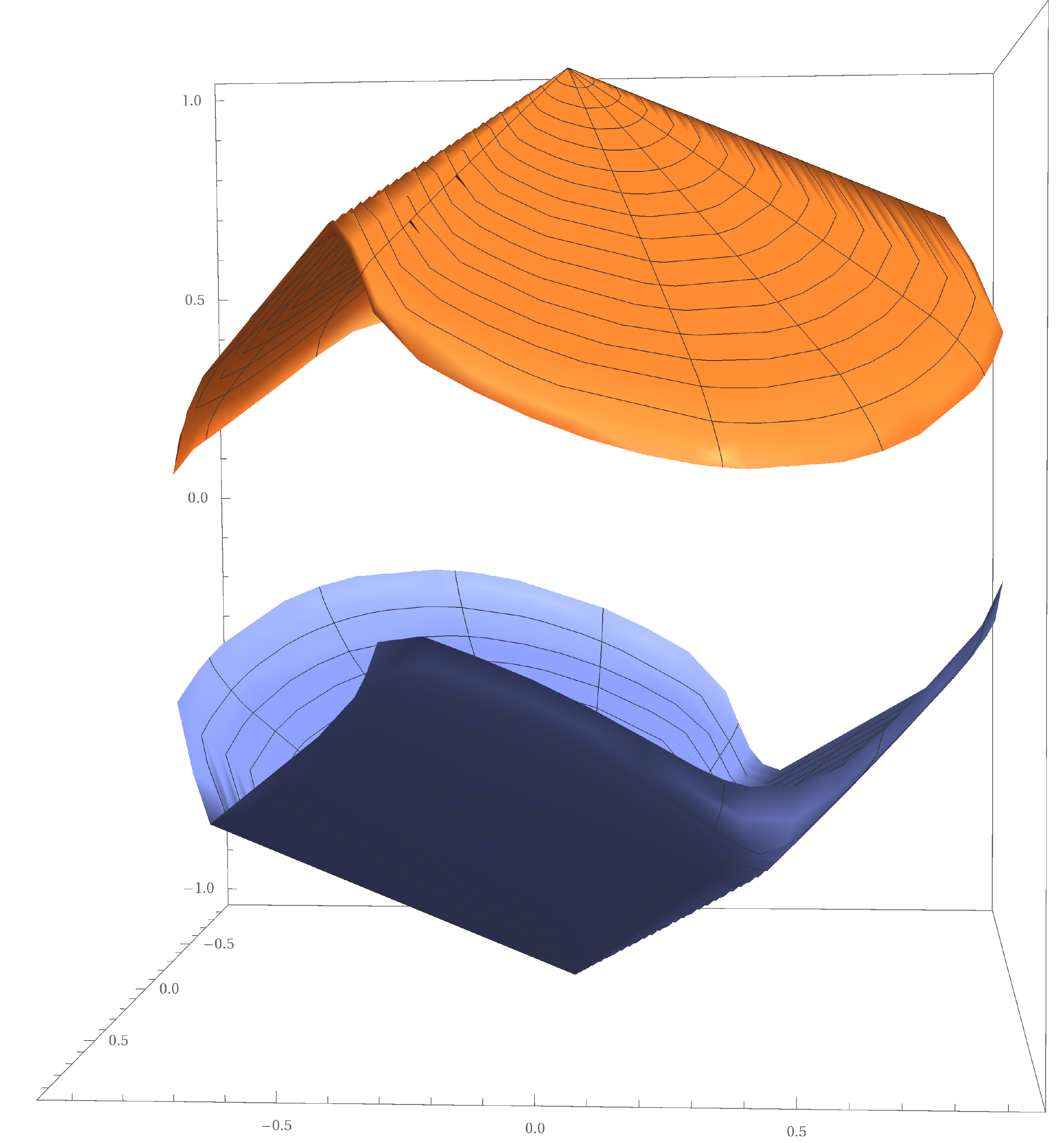}
\end{center}
\caption{The lozenge network (left) and a plot (right) of the functions 
$\cS\ni\xi\mapsto\Lambda_\pm(\xi)$ in thermal equilibrium (for the purpose of this 
representation, the set $\cS$ has been mapped to the open unit disk).}
\label{FigLozenge}
\end{figure}

As a first example of numerical exploitation of our scheme, we investigate some
properties of the Lozenge network of Figure~\ref{FigLozenge}. With $|\cI|=4$ and
$|\pI|=3$, the parameters of the model are given by
\[
\kappa^2=\begin{bmatrix}
1&0&\varepsilon&\varepsilon\\
0&1&\varepsilon&\varepsilon\\
\varepsilon&\varepsilon&1&0\\
\varepsilon&\varepsilon&0&1
\end{bmatrix},\qquad
\varepsilon=\frac1{2\sqrt{2}},\qquad
\gamma_1=\gamma_2=\gamma_3 =1.
\]
Consider first the case of thermal equilibrium: $[1:1:1]$.  The mean heat fluxes
vanish, $\bar\varphi=0$. From the right pane of figure~\ref{FigLozenge}, which
shows the functions $\cS\ni\xi\to\Lambda_\pm(\xi)$, one infers that
Condition~\textbf{(R)} is verified so that, by Theorem~\ref{theoglobalLDP}, the
global LDP~\eqref{LDFagain} holds with a rate function $I$ satisfying the
FR~\eqref{IFR} on $\cL^\perp$.

\begin{figure}[h]
\begin{center}
\includegraphics[scale=0.5]{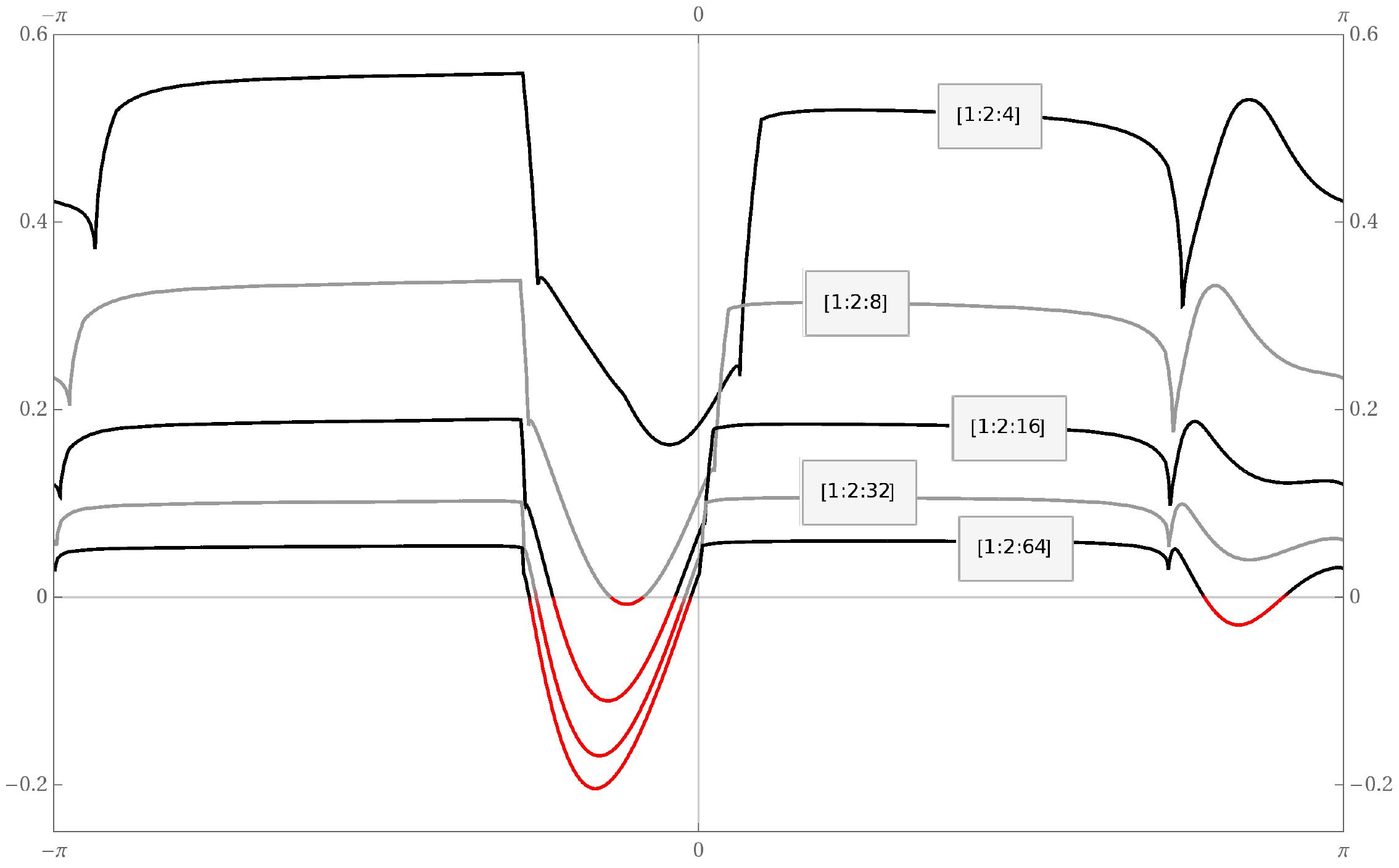}
\end{center}
\caption{Plot of the spectral gap $\partial\cS\ni\xi\mapsto\Lambda_+(\xi)-\Lambda_-(\xi)$
of the lozenge network for different temperature ratios 
(here, the set $\partial\cS$ has been mapped to a circle and the polar angle $0$ corresponds to 
the direction $\vartheta^{-1}$).}
\label{GapLozenge}
\end{figure}

\begin{figure}[!h]
\begin{center}
\includegraphics[scale=0.35]{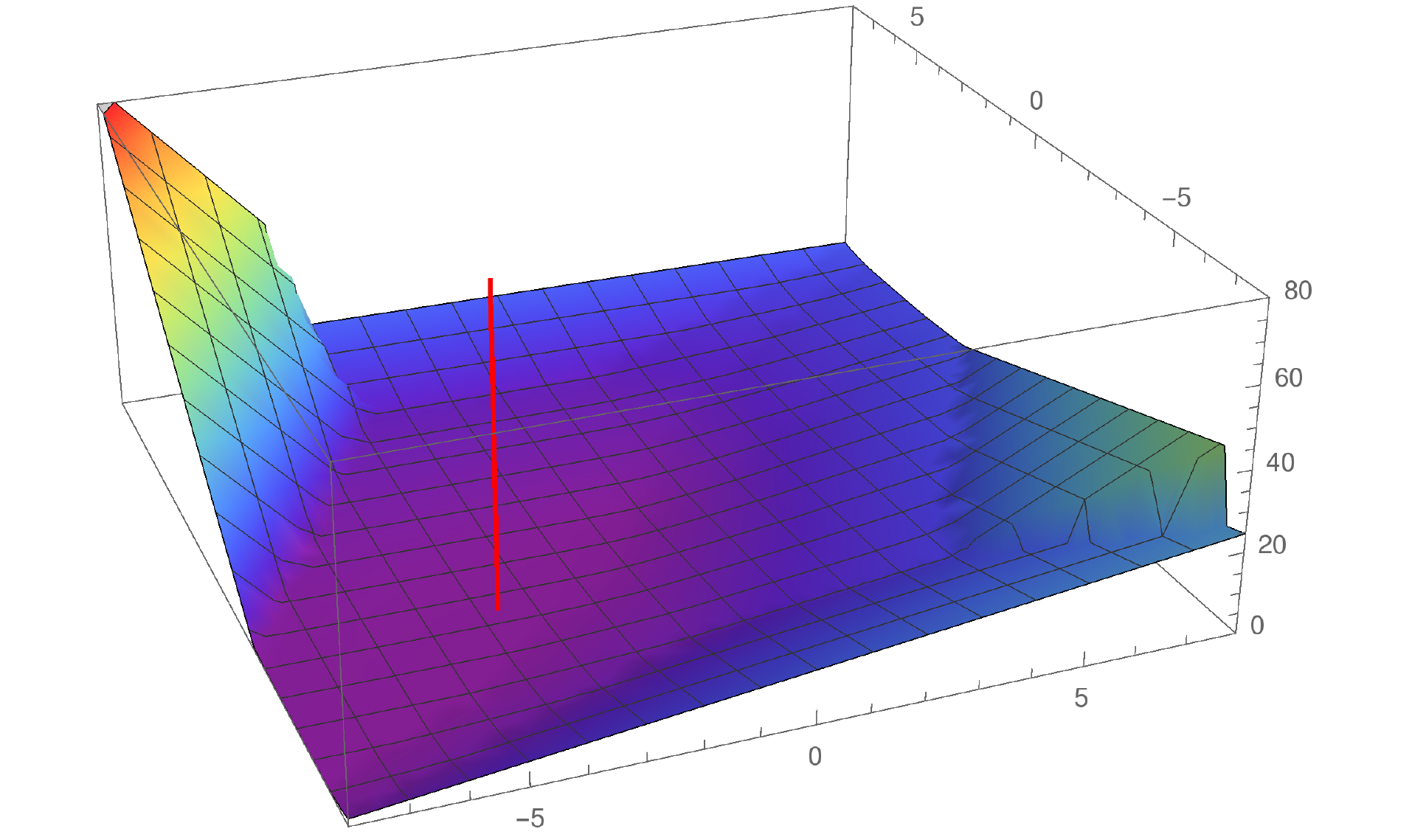}
\includegraphics[scale=0.35]{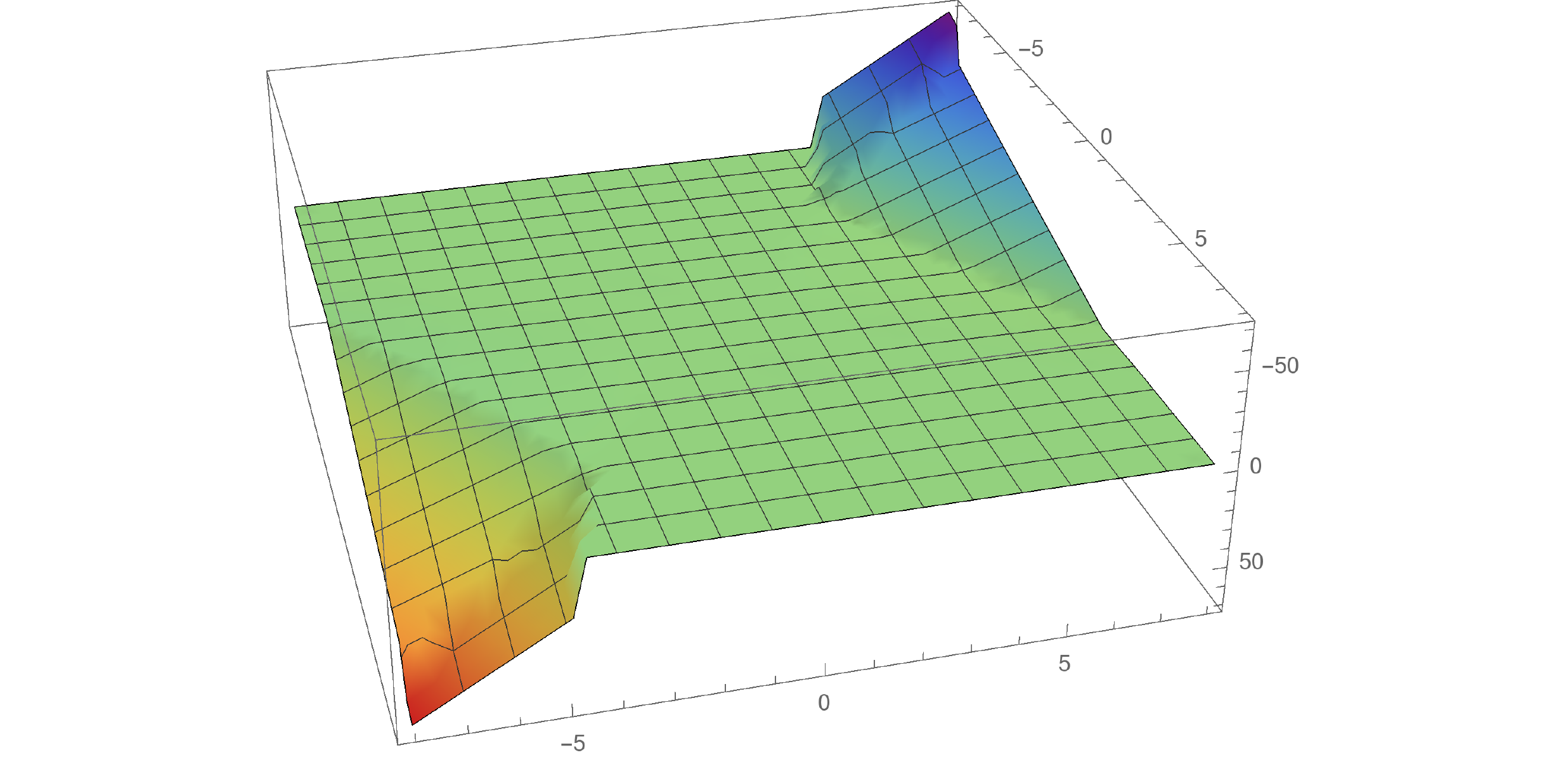}
\end{center}
\caption{The rate function $I$ (left, the vertical line denotes the position of the average 
current~$\bar{\varphi}$) and the anomalous fluctuation function~$\Delta$ (right) for the lozenge 
network (see the main text for details).}
\label{RateCumulLozenge}
\end{figure}

By continuity, Condition~\textbf{(R)} persists sufficiently near thermal
equilibrium so that the same conclusions hold there. Figure~\ref{GapLozenge}
shows the ``spectral gap'' $\Lambda_+-\Lambda_-$ on the boundary of the set
$\cS$ for different temperature ratios. It appears that this gap eventually
closes (i.e., takes non-positive values) when the temperature differences become
large. We conclude that the FR~\eqref{IFR} breaks down in this regime. This is
illustrated on Figure~\ref{RateCumulLozenge} where the rate function $I$ and the
anomalous fluctuation function
$\Delta(\varphi)=I(\varphi)-I(-\varphi)-\vartheta^{-1}\cdot\varphi$ are plotted
for the temperature ratios $[1:2:64]$.

\subsection{A triangular network}

\begin{figure}[h]
\begin{center}
\includegraphics[scale=0.7]{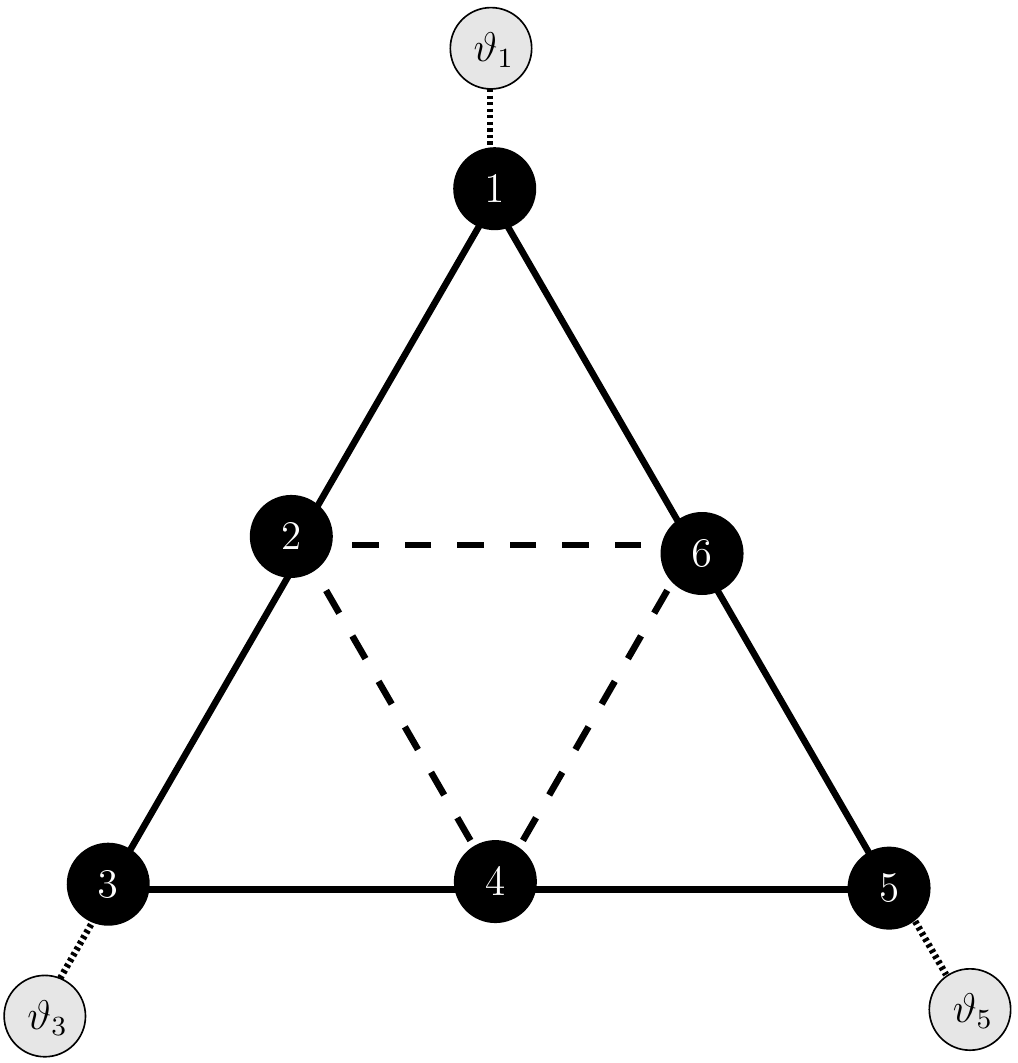}
\includegraphics[scale=0.45]{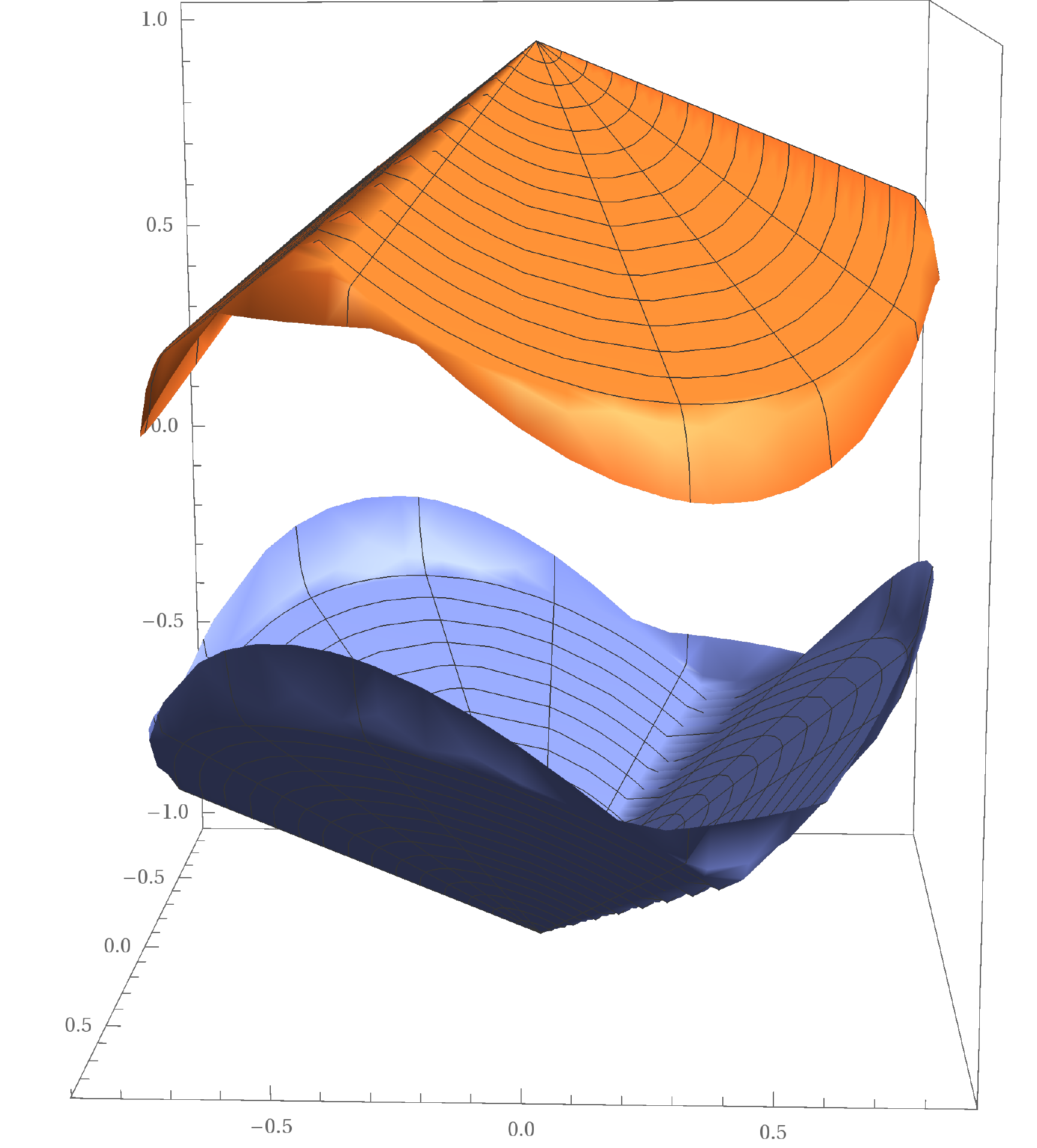}
\end{center}
\caption{A triangular network (left ) and a plot (right) of the functions $\cS\ni\xi\mapsto\Lambda_+\pm(\xi)$ in thermal equilibrium.}
\label{FigTriangle}
\end{figure}

Our second example is the triangular network already considered in~\cite{JPS}
and illustrated on the left pane of Figure~\ref{FigTriangle}. Here we have
$|\cI|=6$, $|\pI|=3$ and the parameters are
\[
\kappa^2=\begin{bmatrix}
1/2&a&0&0&0&a\\
a&1/2&a&b&0&b\\
0&a&1/2&a&0&0\\
0&b&a&1/2&a&b\\
0&0&0&a&1/2&a\\
a&b&0&b&a&1/2
\end{bmatrix},
\qquad
a = \frac{1}{2\sqrt{2}}, \qquad b=\frac{1}{4}, \qquad \gamma_1 = \gamma_3 =\gamma_5 =1.
\]
In thermal equilibrium one finds that $\cS$ is the disk of radius $\sqrt{3}/2$
centered at $0$ on $\cL^\perp$. The spectral gap $\Lambda_+-\Lambda_-$ is open,
as seen on the right pane of Figure~\ref{FigTriangle}.  Hence, here again,
Theorem~\ref{theoglobalLDP} applies: the global LDP~\eqref{LDFagain}  and the
FR~\eqref{IFR} hold on $\cL^\perp$ near equilibrium.
\begin{figure}[!h]
\begin{center}
\includegraphics[scale=0.4]{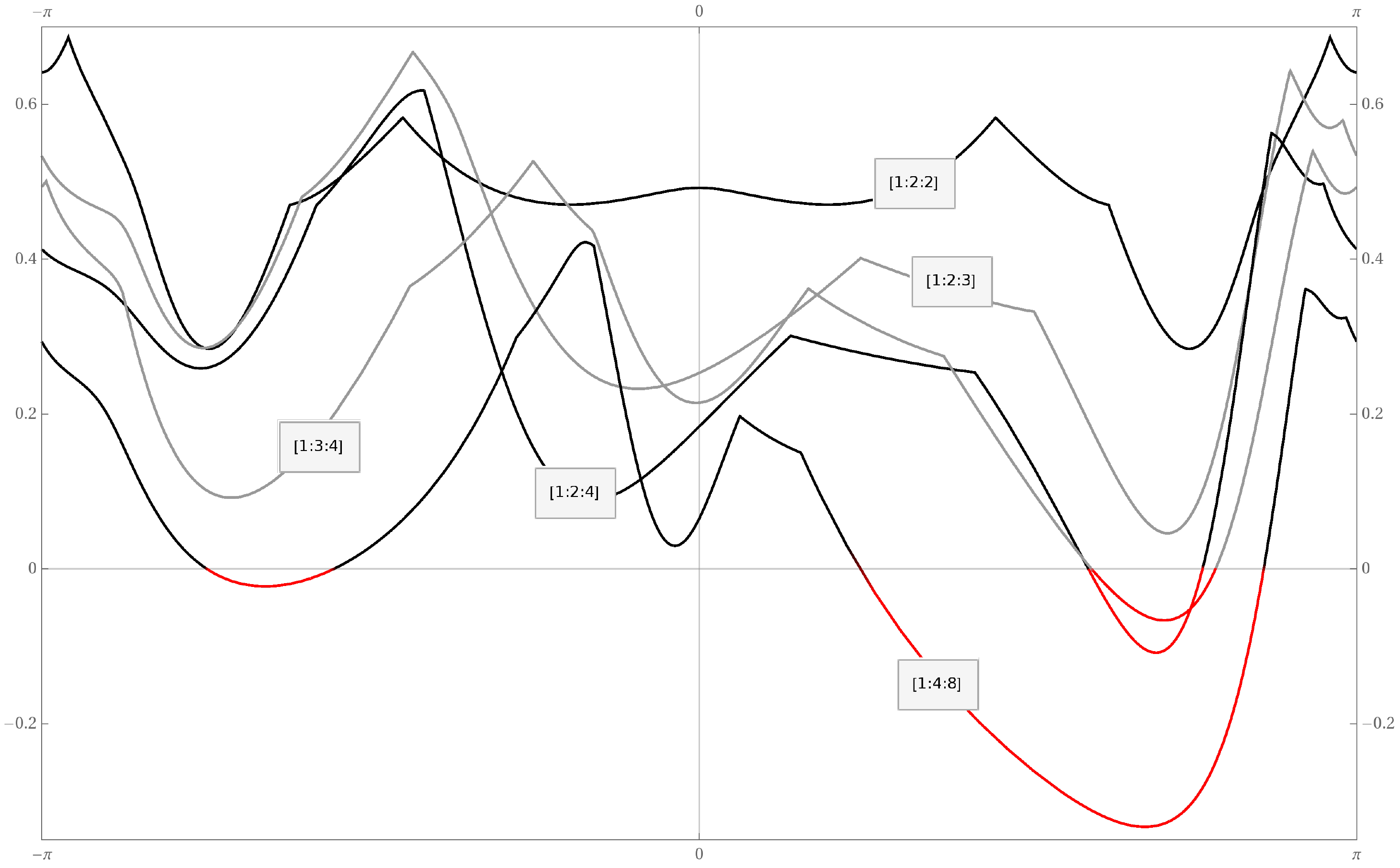}
\end{center}
\caption{The spectral gap $\Lambda_+-\Lambda_-$ of the triangular network on the boundary 
$\partial\cS$ for different temperature ratios.}
\label{GapTriangle}
\end{figure}
The spectral gap on the boundary $\partial\cS$ is plotted in Figure~\ref{GapTriangle} for various 
temperature ratios $[\vartheta_1:\vartheta_2:\vartheta_3]$. One observes a similar behavior as 
in our first example.

\subsection{A heat pump network}

\begin{figure}[!h]
\begin{center}
\includegraphics[scale=0.7]{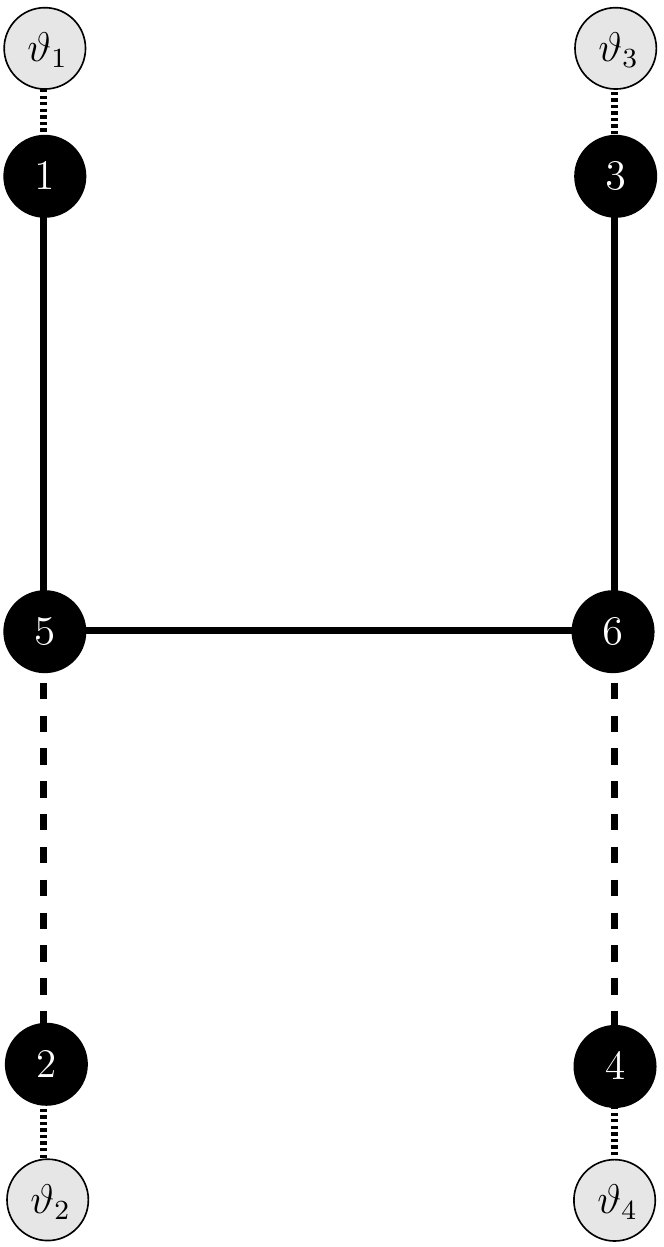}
\hskip 1cm
\raisebox{1.5cm}{\includegraphics[scale=0.43]{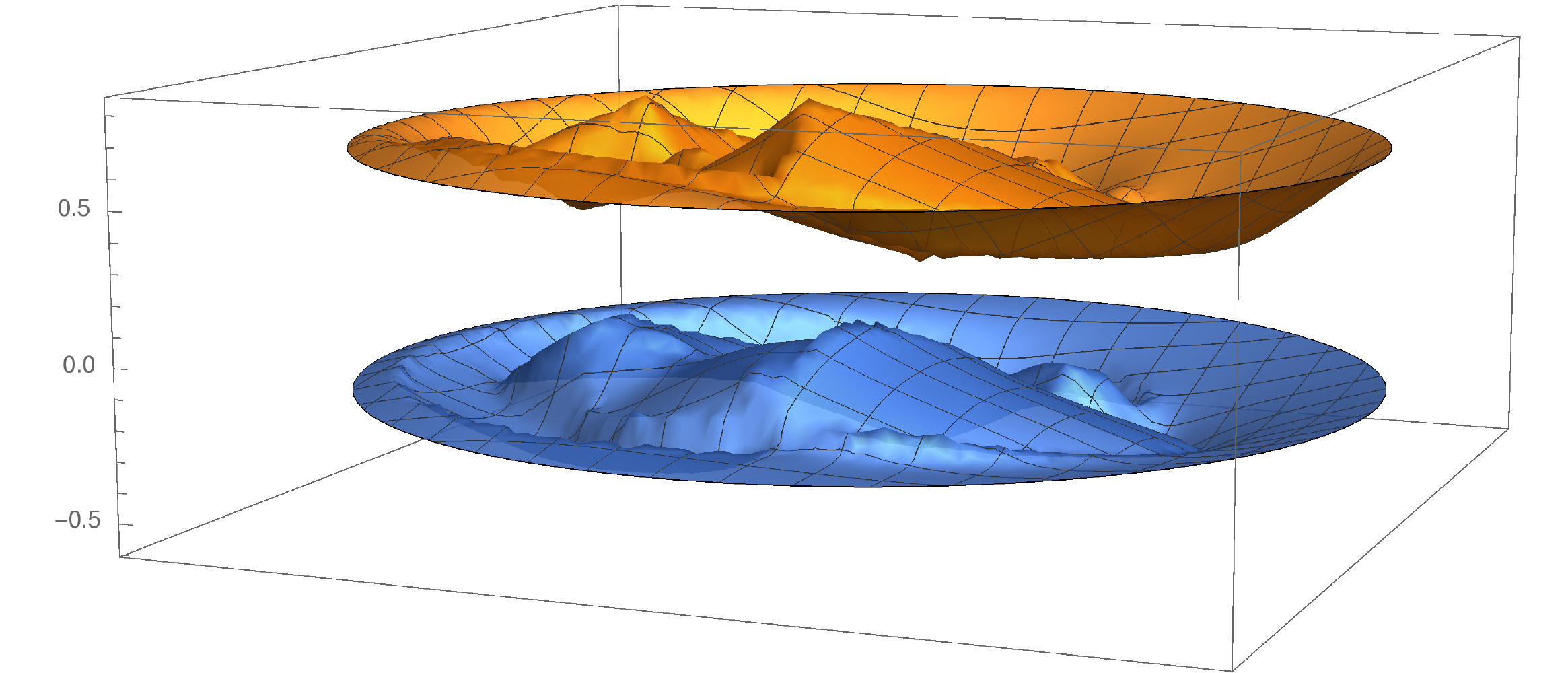}}
\end{center}
\caption{The heat pump network (left) and a plot of the functions
$\partial\cS\ni\xi\mapsto\Lambda_\pm(\xi)$ for the temperature ratios $[10:3.6:7:6.8]$. }
\label{FigHeatPump}
\end{figure}
\begin{figure}[!h]
\begin{center}
\includegraphics[scale=0.45]{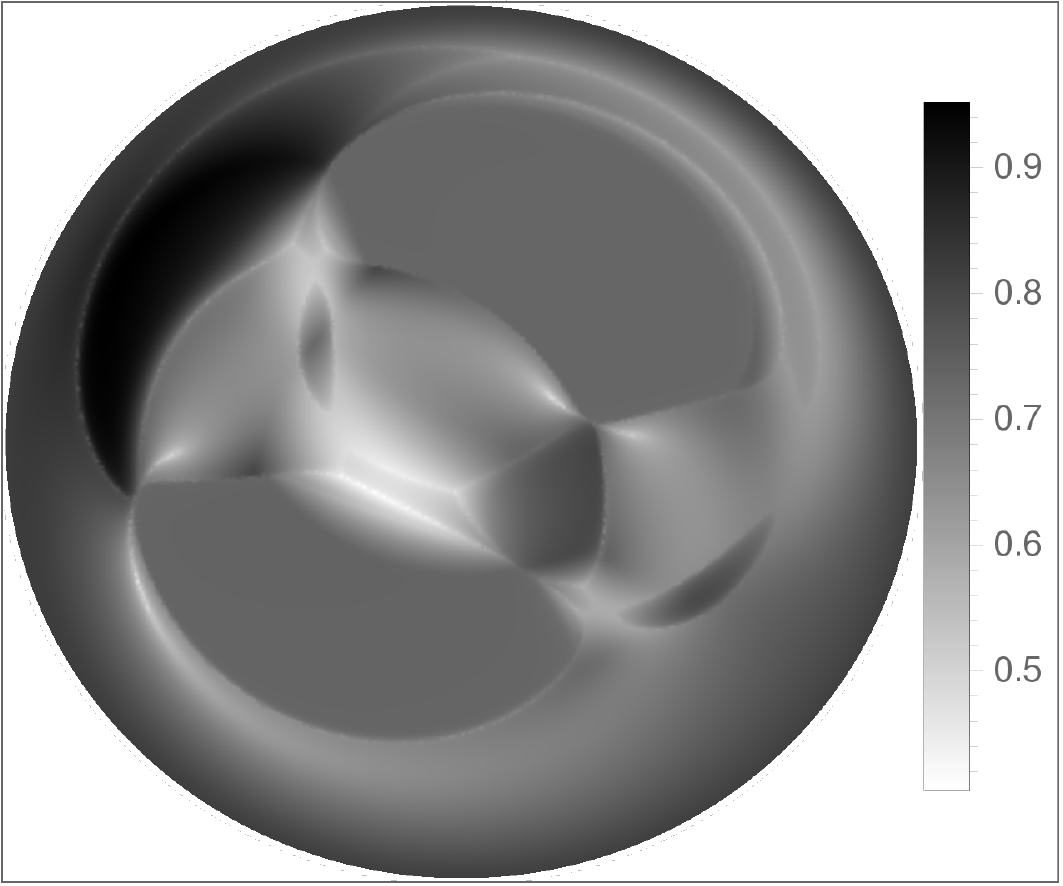}
\hskip 0.5cm
\includegraphics[scale=0.45]{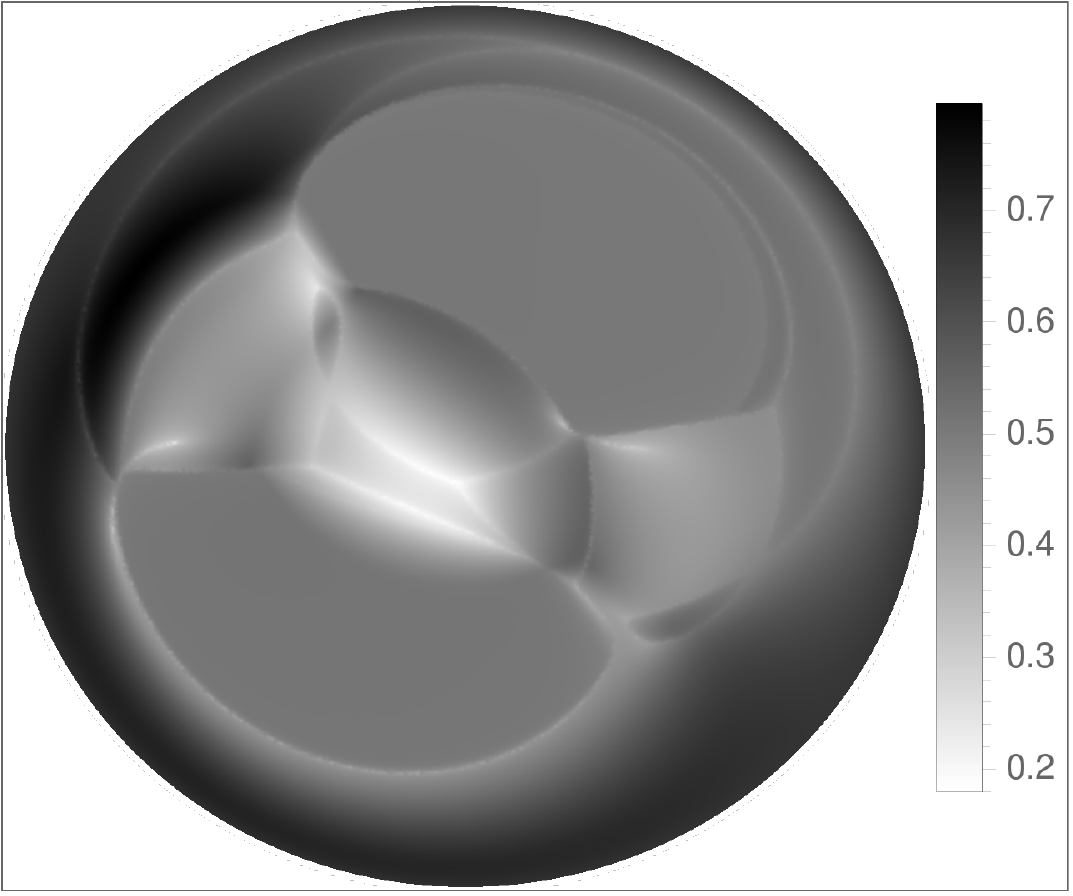}
\hskip 0.5cm
\includegraphics[scale=0.45]{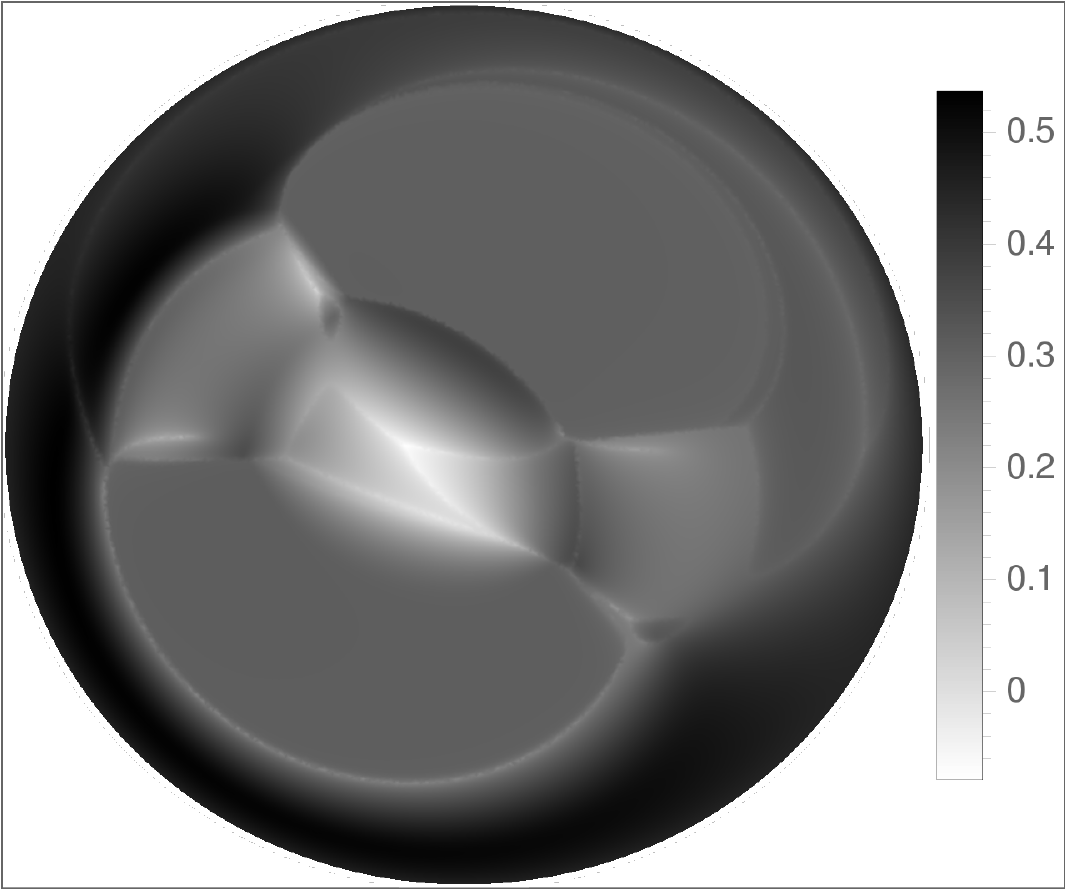}
\end{center}
\caption{Density plots of the spectral gap $\Lambda_+-\Lambda_-$ as a function of $\xi\in\partial\cS$ for the temperature ratios
$[10:3.6:7:6.8]$, $[20:3.6:7:6.8]$ and $[40:3.6:7:6.8]$.}
\label{FigGapHeatPump}
\end{figure}
Our last example is the heat pump network of~\cite{EZ}, see
Figure~\ref{FigHeatPump}. With $|\cI|=6$ and $|\pI|=4$, the parameters:
\[
\kappa^2=\begin{bmatrix}
1-a&0&0&0&a&0\\
0&1-b&0&0&b&0\\
0&0&1-a&0&0&a\\
0&0&0&1-b&0&b\\
a&b&0&0&1-2a-b&b\\
0&0&a&b&a&1-2a-b
\end{bmatrix}
\begin{tabular}{cccc}
&$a=-40,$&$b=-20,$&\\
&&&\\
\multicolumn{4}{c}{$\gamma_1=\gamma_2=\gamma_3=\gamma_4=1,$}\\
&&&\\
$\vartheta_1=10,$&$\vartheta_2=3.6,$&$\vartheta_3=7,$&$\vartheta_4=6.8,$
\end{tabular}
\]
were chosen in~\cite{EZ}  in such a way that the mean steady heat current between the vertices 
5 and 6 vanishes while the heat flows from the hot reservoir to the cold one on the left side, and 
from the cold to the hot one on the right side. Thus, the right side of the device acts as a heat 
pump. On the right pane of  Figure~\ref{FigHeatPump} we plot the two functions 
$\Lambda_\pm$ on the boundary $\partial\cS$.\footnote{ %
In Figures~\ref{FigHeatPump} and~\ref{FigGapHeatPump} the set $\partial\cS$ is
mapped to the closed unit disk by first mapping $\partial\cS$ to the unit sphere and then
mapping the point with spherical coordinates
$(\varphi,\theta)\in[0,2\pi]\times[0,\pi]$ on this sphere to the point
$\frac\theta\pi(\cos\varphi,\sin\varphi)$ of the plane.} %
Condition~\textbf{(R)} is satisfied, so that the global large 
deviation principle~\eqref{LDFagain}  and the FR~\eqref{IFR} hold on $\cL^\perp$
 near this non-equilibrium heat pump regime.

As shown in Figure~\ref{FigGapHeatPump}, here again the spectral gap closes as the temperatures differences increase.

\section{Proofs}
\label{SCT Proofs}

\subsection{Proof of Proposition~\ref{propgxi}}

In order to prove the proposition, we shall need the following two Lemmas.
\begin{lemm}
Let $A$ and $B$ be linear operators on a finite dimensional Hilbert space and
$\C=V_-\cup\ell\cup V_+$ a partition of the plane into two open half-planes
$V_{\pm}$ and a separating line $\ell$. Assume that
$(\sp(A)\cup\sp(B))\cap\ell=\emptyset$ and that $A$ and $B$ have the same number
of repeated eigenvalues in $V_+$. Denote by $A_\pm$ and $B_\pm$ the parts of $A$
and $B$ corresponding to their spectra in $V_\pm$. Let $K$ be a compact set such
that $(\sp(A)\cup\sp(B))\cap V_+\subset K\subset V_+$. Then the following holds:
\begin{enumerate}[(1)]
\item The function $f(z)=\log\det\left((z-A)^{-1}(z-B)\right)$ is analytic in
$V_+\setminus K$.
\item For any Jordan curve $\gamma$ in $V_+\setminus K$ ``enclosing'' $K$
\begin{equation}
\oint_\gamma f(z)\frac{\d z}{2\pi\i}=-\oint_\gamma zf'(z)\frac{\dz}{2\pi\i}
=\tr(A_+-B_+).
\label{eqff'}
\end{equation}
\end{enumerate}
\end{lemm}
\textbf{Proof.} {\bf (1)} By assumption we can enumerate the repeated eigenvalues of $A$ and
$B$ in such a way that
\[
\sp(A)=\{\lambda_j^+\,|\,j\in J\}\cup\{\lambda_i^-\,|\,i\in I\},\qquad
\sp(B)=\{\mu_j^+\,|\,j\in J\}\cup\{\mu_i^-\,|\,i\in I\}
\]
with
\[
\lambda_j^+,\mu_j^+\in K\subset V_+,\qquad
\lambda_j^-,\mu_j^-\in V_-.
\]
In terms of these eigenvalues, we have
\[
\det\left((z-A)^{-1}(z-B)\right)=
\left(\prod_{j\in J}\frac{z-\mu_j^+}{z-\lambda_j^+}\right)
\left(\prod_{i\in I}\frac{z-\mu_i^-}{z-\lambda_i^-}\right).
\]
The Möbius transformation $z\mapsto\frac{z-b}{z-a}$ maps the interior of the
complement of any open neighborhood of the line segment joining $a$ to $b$ to a
simply connected open subset of $\C\setminus\{0\}$. It follows that the function
$\log\frac{z-b}{z-a}$ is analytic on the complement of any neighborhood of the
segment joining $a$ to $b$. Thus, the functions
$z\mapsto\log\frac{z-\mu_j^+}{z-\lambda_j^+}$ and
$z\mapsto\log\frac{z-\mu_i^-}{z-\lambda_i^-}$ are analytic in $V_+\setminus K$
and so is $f(z)$.

{\bf (2)} The first identity in~\eqref{eqff'} now follows from integration by parts.
Finally, noticing that
\[
f'(z)=\sum_{j\in J}\left(\frac1{z-\mu_j^+}-\frac1{z-\lambda_j^+}\right)
+\sum_{i\in I}\left(\frac1{z-\mu_i^-}-\frac1{z-\lambda_i^-}\right)
=\tr(z-B)^{-1}-\tr(z-A)^{-1},
\]
the second identity follows from the Riesz formula
\[
-\oint_\gamma zf'(z)\frac{\d z}{2\pi\i}
=-\tr\left(\oint_\gamma z(z-B)^{-1}\frac{\d z}{2\pi\i}\right)
+\tr\left(\oint_\gamma z(z-A)^{-1}\frac{\d z}{2\pi\i}\right)
=\tr(A_+)-\tr (B_+).
\]
\hfill$\square$

\begin{lemm}\label{lemma2}
Let $A$ and $B$ be as in the previous Lemma, where $\ell$ is the imaginary axis and
$V_+$ the right half-plane. Then
\[
\int_{-\infty}^{+\infty}\log\det\left((\i\omega -A)^{-1}(\i\omega-B)\right)
\frac{\d\omega}{4\pi}=\frac14\left(\tr(B_+-B_-)-\tr(A_+-A_-)\right).
\]
In particular, if the spectra of $A$ and $B$ are symmetric w.r.t.\;$\ell$, then
$\tr(A_-)=-\overline{\tr(A_+)}$ and similarly for $B$, so
\begin{eqnarray*}
\int_{-\infty}^{+\infty}\log\det\left((\i\omega-A)^{-1}(\i\omega-B)\right)
\frac{\d\omega}{4\pi}
&=&\frac12\Re(\tr(B_+-A_+))\\
&=&\frac14\sum_{\lambda\in\sp(B)}|\Re(\lambda)|m_\lambda-\frac14\sum_{\lambda\in\sp(A)}|\Re (\lambda)|m_\lambda,
\end{eqnarray*}
where $m_\lambda$ denotes the algebraic multiplicity of the eigenvalue $\lambda$.
\end{lemm}
\textbf{Proof.} Denote by $\gamma_R$ the positively oriented boundary of the
intersection of the disk of radius $R$ centered at $0$ with the right
half-plane. Applying the previous Lemma and observing that $f$ is analytic in a
neighborhood of $\ell$ we get, for $R$ large enough,
\[
\tr(A_+)-\tr(B_+)=\oint_{\gamma_R}f(z)\frac{\d z}{2\pi\i}
=-\int_{-R}^R f(\i\omega)\frac{\d\omega}{2\pi}
+\int_{-\frac{\pi}{2}}^{\frac{\pi}{2}}f(R\e^{\i\varphi})R\e^{\i\varphi}
\frac{\d\varphi}{2\pi}.
\]
To evaluate the second integral on the right-hand side we note that, as $R\to\infty$,
\[
\frac{R\e^{\i\varphi}-b}{R\e^{\i\varphi}-a}
=(1-bR^{-1}\e^{-\i\varphi})(1+aR^{-1}\e^{-\i\varphi}+O(R^{-2}))
=1+(a-b)R^{-1}\e^{-\i\varphi}+O(R^{-2})
\]
so
\[
\log\frac{R\e^{\i\varphi}-a}{R\e^{\i\varphi}-b}
=(a-b){R^{-1}\e^{-\i\varphi}}+O(R^{-2}),
\]
and
\[
f(R\e^{\i\varphi})=R^{-1}\e^{-i\varphi}\tr(A-B)+O(R^{-2}).
\]
It follows that
\[
\lim_{R\to\infty}\int_{-\frac{\pi}{2}}^{\frac{\pi}{2}}
f(R\e^{\i\varphi})R\e^{\i\varphi}\frac{\d\varphi}{2\pi}
=\frac12\tr(A-B),
\]
and we conclude that
\begin{align*}
\int_{-\infty}^{+\infty}\log\det\left((\i\omega-A)^{-1}(\i\omega-B)\right)
\frac{\d\omega}{4\pi}
=&\frac14\tr(A-B)-\frac12(\tr(A_+)-\tr(B_+))\\
=&-\frac14\tr(A_+)+\frac14\tr(A_-)+\frac14\tr(B_+)-\frac14\tr(B_-).
\end{align*}
The last two statements follow from elementary calculations.
\hfill $\square$

\bigskip
We now turn to the proof of Proposition~\ref{propgxi}

\textbf{(1)} Set $R(\omega)=\vartheta^{-1}Q^\ast(A+\i\omega)^{-1}Q$. Assumption
{\bf (C)} implies that $A$ is stable (see~\eqref{structuralconstraints1}) so
that the map $\R\ni\omega\mapsto R(\omega)\in L(\Xi)$ is continuous. Using the
identities
\[
\Omega=A+\frac12Q\vartheta^{-1}Q^\ast=-A^\ast-\frac12Q\vartheta^{-1}Q^\ast
\]
a simple calculation yields that for any $\txi\rhd\xi\in\Xi$ one has
\begin{equation}
E_\xi(\omega)=Q^\ast(A^\ast-\i\omega)^{-1}[\Omega,\txi]
(A+\i\omega)^{-1}Q
=-\zeta R(\omega)-R(\omega)^\ast\zeta-R(\omega)^\ast\zeta R(\omega),
\label{Eq:FdecayForm}
\end{equation}
with $\zeta=\vartheta^{1/2}\xi\vartheta^{1/2}$.
All the stated properties immediately follow.

\bigskip
\textbf{(2)} It will be convenient to introduce $U(\omega)=I+R(\omega)$ and to rescale $\xi$
by setting $\zeta=(\xi_i\vartheta_i)_{i\in\pI}$. With this change of variable
\[
I-E_\xi(\omega)=I-F_\zeta(\omega)=I-\zeta+U(\omega)^\ast\zeta U(\omega),
\]
and
\[
\cD=\bigcap_{\omega\in\R}\{\zeta\in\Xi\,|\,I-F_\zeta(\omega)>0\},\qquad
\cD_0=\{\zeta\in\Xi\,|\,0<\zeta_i<1,i\in\pI\}.
\]
It immediately follows that $\cD_0\subset\cD$. Moreover, since $I-F_\zeta(\omega)$
is an affine function of $\zeta$, $\cD$ is convex.

Using~\eqref{structuralconstraints1}, elementary calculations
(see~\cite[Section~5.5]{JPS}) show that, for any $\omega\in\R$,
$U(\omega)^{-1}=U(-\omega)$ and $|\det(U(\omega))|=1$. The first relation allows
us to derive
\begin{equation}
I-F_\zeta(\omega)=U(\omega)^\ast(U(-\omega)^\ast(I-\zeta)U(-\omega)-(I-\zeta)+I)U(\omega)
=U(\omega)^\ast(I-F_{I-\zeta}(-\omega))U(\omega),
\label{Eq:etaSym}
\end{equation}
and the second one yields that $\zeta\in\cD\Longleftrightarrow I-\zeta\in\cD$,
which shows that $\cD$ is centrally symmetric around the point $I/2$. To show
that it is open, we now argue that its complement $\cD^\mathrm{c}$ is closed.
Indeed, $\xi\in\cD^\mathrm{c}$ iff there exists $\omega\in\R$ such that
$E_\xi(\omega)$ has an eigenvalue $\lambda\ge1$. Thus, if $\xi_n$ is a sequence
in $\cD^\mathrm{c}$ which converges to $\xi$, there exist $\omega_n\in\R$,
$\lambda_n\ge1$ and unit vectors $u_n$ such that
$E_{\xi_n}(\omega_n)u_n=\lambda_n u_n$. Given the fact that
\[
1\le\lambda_n\le\|E_{\xi_n}(\omega_n)\|\le C(1+|\omega_n|)^{-2}\le C<\infty
\]
one concludes that $|\omega_n|\le C^{1/2}-1$ and $\lambda_n\in[1,C]$. Thus the
sequence $(\xi_n,\omega_n,\lambda_n,u_n)$ has a convergent subsequence with
limit $(\xi,\omega,\lambda,u)$ satisfying $E_{\xi}(\omega)u=\lambda u$,
$\lambda\ge1$ and $u\not=0$. Hence, $\xi\in\cD^\mathrm{c}$ and consequently
$\cD^\mathrm{c}$ is closed.

If the map $\omega\mapsto E_\eta(\omega)$ vanishes identically on $\R$, then it follows
from the identity $I-E_{\xi+\lambda\eta}(\omega)=I-E_{\xi}(\omega)$ that
$\cD+\lambda\eta\subset\cD$ for all $\lambda\in\R$. Reciprocally, if the later
condition holds, then for all $\xi\in\cD$ and all $\lambda>0$, one has
\[
-\frac1\lambda(I-E_\xi(\omega))<E_\eta(\omega)<\frac1\lambda(I-E_\xi(\omega)).
\]
Letting $\lambda\to\infty$ we deduce that $E_\eta(\omega)$ vanishes identically.
Note that the later condition is satisfied for any $\eta$ such that there exists
$\teta\rhd\eta$ with $\Sigma_\teta=0$. This is in particular the case for
$\eta=\boldsymbol{1}$ and $\teta=I_\Gamma$. Reciprocally, since it follows from
Condition~\textbf{(C)} that the set
$\{(A+\i\omega)^{-1}Qu\,|\omega\in\R,u\in\C^\pI\}$ is total in $\C^{\cI}$, one
concludes that $E_\xi(\omega)=0$ for all $\omega\in\R$ implies $\Sigma_\txi=0$
for any $\txi\rhd\xi$.

\bigskip
\textbf{(3)} Consider first the map
$\cD\ni\eta\mapsto-\log\det(I-F_\eta(\omega))$ for fixed $\omega\in\R$. The
previous discussion clearly implies that it is real analytic. Its convexity
follows from an elementary calculation which yields that
\[
-\sum_{i,j\in\pI}\bar{z}_i\frac{\partial^2\log\det(I-F_\eta(\omega))}
{\partial\eta_i\partial\eta_j}z_j
=\tr\left((I-F_\eta(\omega))^{-1/2}F_z(\omega)^\ast(I-F_\eta(\omega))^{-1}
F_z(\omega)(I-F_\eta(\omega))^{-1/2}\right)\ge0
\]
for $\eta\in\cD$ and $z\in\C^\pI$. From~\eqref{Eq:FdecayForm} one further
deduces that $F_\eta(\omega)=O(\omega^{-2})$ as $|\omega|\to\infty$, locally
uniformly in $\eta\in\cD$. It follows that
\[
f(\eta)=-\int_{-\infty}^{\infty}\log\det(I-F_\eta(\omega))\frac{\d\omega}{4\pi}=g(\xi)
\]
is convex and real analytic on $\cD$. The identity~\eqref{Eq:etaSym} leads to
\[
\det(I-F_\eta(\omega))=\det(I-F_{\boldsymbol{1}-\eta}(-\omega)),
\]
and in particular $\det(I-F_{\boldsymbol{1}}(\omega))=1$. This proves the first
equality in~\eqref{G-Csymmetry}. The second one follows from~\eqref{DlinForm}
and the linearity of the map $\eta\mapsto F_\eta$.

\bigskip
\textbf{(4)}  The second equality in~\eqref{G-Csymmetry} implies that
$\eta\cdot\nabla g(\xi)=0$ for all $\xi\in\cD$ and all $\eta\in\cL$. To
establish the reciprocal property, note that since $\cD$ is open, for any
$\xi\in\cD$ and any $\eta\in\nabla g(\cD)^\perp$ there exists $\epsilon>0$ such
that $\xi+\alpha\eta\in\cD$ for $|\alpha|<\epsilon$. It follows that the
function $\alpha\mapsto g(\xi+\alpha\eta)$ is constant in a real neighborhood of
$0$ and hence extends by analyticity to the constant function on the line
$\xi+\R\eta$. Since, by Part~(7), $g$ is singular on $\partial\cD$, it follows
that $\xi+\R\eta\subset\cD$, i.e., $\eta\in\cL$.

Consequently, $g$ vanishes identically whenever $\cL=\Xi$. In the opposite case,
the calculation in Part~(3) gives that the Hessian of $g$ satisfies
\[
\eta\cdot g''(\xi)\eta=\int_{-\infty}^\infty\tr\left(\left[
(I-E_\xi(\omega))^{-1/2} E_\eta(\omega)(I-E_\xi(\omega))^{-1/2}
\right]^2\right)\,\frac{\d \omega}{4\pi}>0
\]
for non-zero $\eta\not\in\cL$. It follows that the restriction of $g''(\xi)$ to
$\cL^\perp$ is positive definite which implies that the restriction of $g$ to
$\cS$ is strictly convex. To show that the closure of $\cS$ is compact, let us
assume that $\cS$ is unbounded. Since $\cS$ is convex and centrally symmetric
w.r.t.\;the orthogonal projection $\xi_0$ of $(2\vartheta)^{-1}$ onto
$\cL^\perp$, it follows that for some non-vanishing $\xi\in\cL^\perp$ one has
$\xi_0+\lambda\xi\in\cS$ for all $\lambda\in\R$, i.e.,
\[
-\frac{1}{|\lambda|}(I-E_{\xi_0}(\omega))\le E_\xi(\omega)
\le\frac{1}{|\lambda|}(I-E_{\xi_0}(\omega))
\]
for all $\omega\in\R$. Letting $|\lambda|\to\infty$ yields that $\xi\in\cL$
which contradicts the fact that $0\not=\xi\in\cL^\perp$.

\bigskip
\textbf{(5)} We start with some simple consequences of Condition~{\bf (C)}. For
a short introduction to the necessary elementary material, we refer the reader
to~\cite[Section~4]{LR}. Since $A_\xi=A+Q\xi Q^\ast$, the pair $(A_\xi,Q)$ is
controllable for all $\xi$. The relation $A^\ast_\xi=-A_{\vartheta^{-1}-\xi}$
shows that the same is true for the pair $(A^\ast_\xi,Q)$. Thus, one has
\[
\bigcap_{n\geq0}\Ker(Q^\ast A_\xi^n)=\bigcap_{n\geq0}\Ker(Q^\ast A^{*n}_\xi)=\{0\}
\]
for all $\xi$. This implies that if $Q^\ast u=0$ and $(A_\xi-z)u=0$ or
$(A^\ast_\xi-z)u=0$, then $u=0$, i.e., no eigenvector of $A_\xi$ or $A^\ast_\xi$
can live in $\Ker Q^\ast$. Assume now $z\in\sp(A_\xi)$ and let $u\neq0$ be a
corresponding eigenvector. Since
\[
A_\xi+A^\ast_\xi=2Q(\xi-(2\vartheta)^{-1})Q^\ast,
\]
taking the real part of $\langle u,(A_\xi-z)u\rangle=0$ we infer
\[
\langle Q^\ast u,(\xi-(2\vartheta)^{-1})Q^\ast u\rangle=\Re(z)|u|^2.
\]
Thus, controllability of $(A_\xi,Q)$ implies that for
$\pm(\xi-(2\vartheta)^{-1})>0$ one has $\sp(A_\xi)\subset\C_\pm$ and in
particular $\sp(A_\xi)\cap\i\R=\emptyset$. Hence, for $\xi>(2\vartheta)^{-1}$
and $\omega\in\R$, Schur’s complement formula yields
\beq
\det(K_\xi-\i\omega)=|\det(A_\xi+\i\omega)|^2
\det\left(I+r_\xi(\omega)^\ast
\xi(\vartheta^{-1}-\xi)r_\xi(\omega)\right)
\label{Eq:Schur}
\eeq
where we have set
\[
r_\xi(\omega)=Q^\ast(A_\xi+\i\omega)^{-1}Q.
\]
One easily checks that $r_\xi(\omega)=r_0(\omega)(I+\xi r_0(\omega))^{-1}$
from which a simple calculation gives
\[
I+r_\xi(\omega)^\ast
\xi(\vartheta^{-1}-\xi)r_\xi(\omega)
=(I+r_0(\omega)^\ast\xi)^{-1}(I-E_\xi(\omega))(I+\xi r_0(\omega))^{-1}.
\]
Inserting the last identity into the right-hand side of~\eqref{Eq:Schur}
and using the fact that
\[
\det(A_\xi+\i\omega)=\det(A+\i\omega)\det(I+\xi r_0(\omega)),
\]
we obtain
\beq
\det(K_\xi-\i\omega)=|\det(A+\i\omega)|^2\det(I-E_\xi(\omega)).
\label{Eq:KxiForm}
\eeq
Both sides of this identity being polynomials in $\xi$, it extends to
all $\xi\in\Xi$. It follows that
\[
\bigcap_{\omega\in\R}\{\xi\in\Xi\,|\,1\not\in\sp(E_\xi(\omega))\}
=\{\xi\in\Xi\,|\,\sp(K_\xi)\cap\i\R=\emptyset\}.
\]
By continuity of the function $\xi\mapsto \displaystyle \min_{\omega\in\R}\det(I-E_\xi(\omega))$,
$\cD$ is the connected component of the point $\xi=0$ in the left-hand side of
this identity.

\bigskip
\textbf{(6)} For $\xi\in\Xi$, $K_\xi$ is $\R$-linear on the real vector space
$\Gamma\oplus\Gamma$. Thus, its spectrum is symmetric w.r.t.\;the real axis.
Observing that $J K_\xi+K_\xi^\ast J=0$, where $J$ is the unitary operator
\[
J=\begin{bmatrix}
0&I\\
-I&0\\
\end{bmatrix},
\]
we conclude that the spectrum of $K_\xi$ is also symmetric w.r.t.\;the imaginary
axis. Assume now that $\xi\in\cD$. Since the eigenvalues of $K_\xi$ are
continuous functions of $\xi$, $K_\xi$ and $K_0$ have the same number of
repeated eigenvalues in the left/right half-plane. From~\eqref{Eq:KxiForm} we
deduce
\[
g(\xi)=\int_{-\infty}^{+\infty}
\log\det\left((\i\omega-K_\xi)^{-1}(\i\omega-K_0)\right)\frac{\d\omega}{4\pi},
\]
and Lemma~\ref{lemma2} allows us to conclude that
\[
g(\xi)=\frac14\sum_{\lambda\in\sp(K_0)}|\Re(\lambda)|m_\lambda
-\frac14\sum_{\lambda\in\sp(K_\xi)}|\Re(\lambda)|m_\lambda .
\]
Since $\sp(K_0)=\overline{\sp(A)}\cup\sp(-A)$ and $A$ is stable, we have
\[
\sum_{\lambda\in\sp(K_0)}|\Re(\lambda)|m_\lambda
=-2\sum_{\lambda\in\sp(A)}\Re(\lambda)m_\lambda
=-2\Re\tr(A)=\tr(Q\vartheta^{-1}Q^\ast),
\]
and ~\eqref{eqgalg} follows for $\xi\in\cD$. Since the eigenvalues of $K_\xi$
are continuous functions of $\xi\in\Xi$, this relation extends to
$\xi\in\overline{\cD}$. The boundedness of this extension follows from the
translation invariance along $\cL$ and the precompactness of $\cS$.

\bigskip
\textbf{(7)} The idea of the proof is that $\xi_0\in\partial\cD$ iff at least
one eigenvalue of $E_{\xi_0}(\omega)$ reaches its global maximum $1$ at some
$\omega_0\in\R$. Since $E_{\xi_0}(\omega)$ is a real analytic function of
$\omega$, the function $\tr\left((1-E_{\xi_0}(\omega))^{-1}\right)$ has a p\^ole
at $\omega=\omega_0$, and since this function is non-negative the order of this
p\^ole must be even. Consequently,
\[
\int_{\omega_0-\epsilon}^{\omega_0+\epsilon}
\tr\left((1-E_{\xi_0}(\omega))^{-1}\right)\d\omega=+\infty.
\]

For $\xi\in\cD$, a simple calculation and Cauchy--Schwarz inequality yield
\[
|\nabla g(\xi)|\ge\frac{\xi}{|\xi|}\cdot\nabla g(\xi)
=\frac1{|\xi|}\int_{-\infty}^{+\infty}
\tr\left((I-E_\xi(\omega))^{-1}E_\xi(\omega)\right)\frac{\d\omega}{4\pi}.
\]
Since $0\not\in\partial\cD$, it suffices to show that the integral on the
right-hand side diverges to $+\infty$ as $\xi\to\xi_0\in\partial\cD$. Let us fix
$\xi_0\in\partial\cD$ and set $V_\delta=\{\xi\in\cD\,|\,|\xi-\xi_0|<\delta\}$.
Elementary considerations show that for sufficiently small $\delta>0$ and
sufficiently large $M>0$ there exists a constant $C$ such that
\[
\int_{-\infty}^{+\infty}
\tr\left((I-E_\xi(\omega))^{-1}E_\xi(\omega)\right)\frac{\d\omega}{4\pi}
\ge C\left(-1+\int_{-M}^M
\tr\left((I-E_\xi(\omega))^{-1}\right)\d\omega
\right)
\]
for any $\xi\in V_\delta$. Making $\delta$ smaller and $M$ larger if necessary,
we can assume that $J_\xi(\omega)=(I-E_{\xi-\xi_0}(\omega))^{-1/2}$ satisfies
$1/\sqrt2\le J_\xi(\omega)\le\sqrt2$ for $(\omega,\xi)\in[-M,M]\times V_\delta$.
Writing
\[
(I-E_\xi(\omega))^{-1}
=J_\xi(\omega)
(I-J_\xi(\omega)E_{\xi_0}(\omega)J_\xi(\omega))^{-1}J_\xi(\omega)
\]
and observing that this implies, in particular, that
$I-J_\xi(\omega)E_{\xi_0}(\omega)J_\xi(\omega)>0$, we derive
\[
\tr\left((I-E_\xi(\omega))^{-1}\right)
\ge\frac12\tr\left(
(I-J_\xi(\omega)E_{\xi_0}(\omega)J_\xi(\omega))^{-1}\right)>0.
\]
By Fatou's lemma
\[
\liminf_{\cD\ni\xi\to\xi_0}\int_{-M}^M
\tr\left((I-E_\xi(\omega))^{-1}\right)\d\omega
\ge\frac12\int_{-M}^M\tr\left((I-E_{\xi_0}(\omega))^{-1}\right)
\d\omega
\]
and by the above argument, the last integral is $+\infty$.

\bigskip
\textbf{(8)} The existence and uniqueness of the maximal solutions of
Eq.~\eqref{Eq:RiccatiFirst} as well as the stated properties of $D_\xi$ follow
from~\cite[Theorems~7.3.7 and~7.5.1]{LR}, Part (5), and the relation
$D_\xi=A+Q(\xi Q^\ast-Q^\ast X_\xi)$. It further follows from~\eqref{eqgalg} and
the symmetries of $\sp(K_\xi)$ discussed in the proof of Part~(5) that
\[
g(\xi)=\frac14\tr(Q\vartheta^{-1}Q^\ast)+\frac12\tr(D_\xi)=\frac12\tr(D_\xi-D_0)
=-\frac12\tr(Q^\ast(X_\xi-\txi)Q).
\]
The proof of Proposition~\ref{propgxi} is complete.

\subsection{Proof of Proposition~\ref{proplimgxi}}

\subsubsection{Some properties of the algebraic Riccati equations~\eqref{Eq:RiccatiFirst}}

In order to prove  Proposition~\ref{proplimgxi} we shall need some properties of the algebraic
Riccati equation
\begin{equation}
\mathcal{R}_\xi(X)\equiv XBX-XA_{\xi}-A^*_{\xi}X-C_{\xi}=0.
\label{riccatiequation}
\end{equation}
This is the purpose of the following proposition which provides a generalization
of~\cite[Proposition~5.5]{JPS}.  In the sequel, whenever we mention a solution
of~\eqref{riccatiequation}, we always mean a {\sl self-adjoint\/} $X\in
L(\Gamma)$ such that $\mathcal{R}_\xi(X)=0$. We say that such a solution $X$ is
maximal (resp.\;minimal) if any other solution $X'\in L(\Gamma)$ satisfies
$X'\le X$ (resp.\;$X'\ge X$).

\begin{prop}\label{propRiccati}
Assume that Condition {\bf(C)} holds.
\begin{enumerate}[(1)]
\item For $\xi\in\cD$  the Riccati equation~\eqref{riccatiequation} has a unique maximal 
solution $X_\xi$ and a unique minimal solution $-\theta X_{\vartheta^{-1} - \xi}\theta$.
Moreover,  the matrix
\[
D_\xi=A_\xi-BX_\xi
\]
is stable and satisfies
\[
Y_\xi=X_\xi+\theta X_{\vartheta^{-1}-\xi}\theta>0.
\]
\item If $\xi_0\in\partial\cD$ is finite, then the non-tangential limit
\[
X_{\xi_0}=\lim_{\cD\ni\xi\to\xi_0} X_{\xi}
\]
exists and is the maximal solution of the corresponding limiting Riccati equation
$\mathcal{R}_{\xi_0}(X)=0$.

\item The function $\cD\ni\xi\mapsto X_\xi\in L(\Gamma)$ is real analytic and concave.
Moreover, $X_\xi<0$ for $\xi<0$, $X_\xi>0$ for $\xi$ in the convex hull of the set
$\cD_0\cup\{\xi\in\cD\,|\,\xi>\vartheta^{-1}\}$, $X_0=0$ and 
$X_{\vartheta^{-1}}=\theta M^{-1}\theta$.

\item For any $\xi\in\overline{\cD}$  and $\eta\in\cL$ one has
\[
X_{\xi+\eta}=X_\xi+\teta.
\]


\item For $t>0$, set
\[
M_{\xi,t}=\int_0^t\e^{sD_\xi}B\e^{sD^\ast_\xi}\d s>0.
\]
Then, for all $\xi\in\overline{\cD}$ one has
\[
\lim_{t\to\infty}M^{-1}_{\xi ,t}=\inf_{t>0}M^{-1}_{\xi,t}=Y_\xi\geq 0,
\]
and $\ker(Y_\xi)$ is the spectral subspace of $D_\xi$ corresponding to its imaginary eigenvalues.

\item Set $\Delta_{\xi,t}=M^{-1}_{\xi,t}-Y_\xi$. For all $\xi\in\overline{\cD}$, one has
\begin{equation}
\e^{tD^\ast_\xi}M^{-1}_{\xi,t}\e^{tD_\xi}=\theta\Delta_{\vartheta^{-1}-\xi,t}\theta,
\end{equation}
and
\[
\lim_{t\to\infty}\frac{1}{t}\log\det(\Delta_{\xi,t})=4g(\xi)-\tr(Q\vartheta^{-1} Q^\ast).
\]
In particular, for $\xi\in\cD$, $\Delta_{\xi,t}\to0$ exponentially fast as $t\to\infty$.

\item Let $\widetilde{D}_\xi=\theta D_{\vartheta^{-1}-\xi}\theta$. Then
\[
Y_\xi\e^{t\widetilde{D}^\ast_\xi}=\e^{tD^\ast_\xi}Y_\xi
\]
for all $\xi\in\overline{\cD}$ and $t\in\R$.

\item  For $\xi\in\cD$ and $\eta\in\Xi$
\[
\eta\cdot\nabla g(\xi)=\frac12\tr\left(\Sigma_\teta Y_\xi^{-1}\right)
\]

\end{enumerate}
\end{prop}
\textbf{Proof.} We refer to~\cite{LR} for a detailed introduction to algebraic
Riccati equations (see also the Appendix in~\cite{JPS} for a summary of the
necessary basic facts). Our proof is similar to that
of~\cite[Proposition~5.5]{JPS}.  The Hamiltonian matrix $K_{\xi}$ associated to
the Riccati equation~\eqref{riccatiequation} is given by
Eq.~\eqref{Hamiltonianmatrix}. Let $\mathcal{H}$ be the complex Hilbert space
$\C\Xi\oplus\C\Xi$ on which $K_{\xi}$ acts. The operator
\[
\Theta=\left[\begin{array}{cc}
0 & \theta\\
\theta & 0\\
\end{array}\right]
\]
acts unitarily on $\mathcal{H}$. We have already observed in the proof of
Proposition~\ref{propgxi}~(6) that for $\xi\in\Xi$, the spectrum of $K_{\xi}$ is
symmetric w.r.t.\;the real axis and the imaginary axis. The time-reversal
covariance relations
\begin{equation}
\theta A_\xi\theta=A_\xi^\ast=- A_{\vartheta^{-1} - \xi}, \quad
\theta B\theta=B^\ast=B,\quad
\theta C_\xi\theta=C_\xi^\ast=C_\xi=C_{\vartheta^{-1} - \xi},
\label{paramABCXi}
\end{equation}
which follow easily from the definitions of the operators $A_\xi$, $B$, $C_\xi$, further
yield $\Theta K_\xi-K_{\vartheta^{-1} -\xi}\Theta=0$ which implies
\begin{equation}
\sp(K_\xi)=\sp(K_{\vartheta^{-1} - \xi}).
\end{equation}
Let $\mathcal{H}_- (K_\xi)$ be the spectral subspace of $K_\xi$ for the part of its spectrum
in the open left half-plane $\C_-$.

\textbf{(1)} By Proposition~\ref{propgxi}~(5), $\sp(K_\xi)\cap\i\R=\emptyset$
for $\xi\in\cD$ and the existence and uniqueness of the maximal and minimal
solutions of the Riccati equation~\eqref{riccatiequation} follow
from~\cite[Theorems~7.3.7 and~7.5.1]{LR}. The relation between minimal and
maximal solutions is a consequence of the Relations~\eqref{paramABCXi} which
imply that
\[
\mathcal{R}_\xi (\theta X \theta)=\theta\mathcal{R}_{\vartheta^{-1}-\xi}(-X)\theta.
\]
By~\cite[Theorems~7.5.1]{LR},  the maximal solution $X_\xi$ is related to the spectral subspace
$\mathcal{H}_-(K_\xi)$ by
\[
\mathcal{H}_- (K_\xi) =\Ran\left[
\begin{array}{c}
I\\X_\xi
\end{array}
\right],
\]
moreover, $\sp(D_\xi)=\sp(K_\xi)\cap\C_-$.

$Y_\xi=X_\xi+\theta X_{\vartheta^{-1} - \xi}\theta$ is called the gap of
Eq.~\eqref{riccatiequation}. As the difference between its maximal and minimal
solutions, it is obviously non-negative. It further has the remarkable property
that for any solution $X$, $\ker(Y_\xi)$ is the spectral subspace of $A_\xi -
BX$ for the part of its spectrum in $\i\R$ \cite[Theorem~7.5.3]{LR}. Since
$\sp(D_\xi)\subset\C_-$, we must have $Y_\xi>0$.

\bigskip
\textbf{(2)} Let $\xi_0\in\partial\cD$ be finite and $\eta\not=0$ be
non-tangential to $\partial\cD$ at $\xi_0$. Set $\xi_t=\xi_0-t\eta$.
W.l.o.g.\;we may assume that $\xi_1\in\cD$. The function
\[
]0,1]\ni t\mapsto Z_t=X_{\xi_t}+tX'_{\xi_1}[\eta]
\]
is concave and its first derivative vanishes at $t=1$. Hence, it is monotone
non-decreasing. We claim that the set $\{ X_\xi\, |\,\xi\in\cD,|\xi|<r\}$ is
bounded in $L(\Gamma)$ for any finite $r$. It thus follows that
\[
X=\lim_{t\downarrow0}X_{\xi_t}=\lim_{t\downarrow0}Z_t=\inf_{t\in]0,1]}Z_t
\]
exists. By continuity, one has $\mathcal{R}_{\xi_0}(X)=0$ and
$\sp(A_{\xi_0}-BX)\subset\overline{\C}_-$, and it follows
from~\cite[Theorem~7.5.1]{LR} that $X$ is the maximal solution of the limiting
Riccati equation. In particular, the  non-tangential limit exists (i.e., does
not depend on the direction).

To prove our claim, we first derive a bound on $\hat X_\xi=Q^\ast X_\xi Q$.
Using $(Q^\ast X_\xi Q)^2\le\|Q\|^2 Q^\ast X_\xi^2Q$, one easily deduces
from~\eqref{riccatiequation} and Cauchy-Schwarz inequality
\[
\tr(\hat X_\xi^2)
\le\|Q\|^2
\tr(Q^\ast X_\xi^2Q)
=\|Q\|^2\left(\tr(C_\xi)+2\tr(\hat X_\xi(\xi-(2\vartheta)^{-1}))\right)
\le b_\xi+a_\xi\tr(\hat X_\xi^2)^{1/2},
\]
where $a_\xi$ and $b_\xi$ are locally bounded functions of $\xi$.
Solving the resulting quadratic inequality yields that  $\tr(\hat X_\xi^2)$, and hence
$\tr(Q^\ast X_\xi^2Q)$ are locally bounded as functions of $\xi$.
Rewriting~\eqref{riccatiequation} as the Lyapunov equation
\[
X_\xi A+A^\ast X_\xi=F_\xi\equiv X_\xi BX_\xi-X_\xi Q\xi Q^\ast-Q\xi Q^\ast X_\xi-C_\xi,
\]
and using the fact that $A$ is stable, we get
\[
X_\xi=-\int_0^\infty\e^{tA^\ast}F_\xi\e^{tA}\d t.
\]
It follows that for any $T\in L(\Gamma)$
\[
|\tr(TX_\xi)|\le\int_0^\infty\left|\tr\left(\e^{tA}T\e^{tA^\ast}F_\xi\right)\right|\d t,
\]
from which one easily concludes that $\|X_\xi\|$ is locally bounded.

\bigskip
\textbf{(3)} The spectral projection of $K_{\xi}$ for the part of its spectrum in $\C_+$ 
can be written as
\[
P_\xi=\left[\begin{array}{c}I\\X_\xi\end{array}\right]
Y_\xi^{-1}
\left[\begin{array}{cc}\theta X_{\vartheta^{-1}-\xi}\theta &I\end{array}\right]
=\left[\begin{array}{cc}
I- Y_{\xi}^{-1}X_{\xi} & Y_{\xi}^{-1}\\
X_{\xi}(I- Y_{\xi}^{-1}X_{\xi}) & X_{\xi}Y_{\xi}^{-1}
\end{array}\right].
\]
Since $\cD\ni\xi\mapsto P_\xi$ is real analytic by regular perturbation theory,
$Y_\xi^{-1}$ and $X_\xi Y_\xi^{-1}$ are real analytic function of $\xi\in\cD$.
The same holds for $Y_\xi$ and $X_\xi=X_\xi Y_\xi^{-1} Y_\xi$.

Invoking the implicit function theorem and using the stability of $D_\xi$, one easily
computes derivatives of the map $\cD\ni\xi\mapsto X_\xi$. The first
derivative is the linear map
\begin{equation}
\Xi\ni\eta\mapsto X'_\xi[\eta]
=\teta-\int_0^\infty\e^{t D^\ast_\xi}\Sigma_\teta\e^{t D_\xi}\d t,
\label{Eq:Xprime}
\end{equation}
where, as usual, we identify $\eta\in\Xi$ with the corresponding diagonal matrix in
$L(\Xi)$ and $\teta\rhd\eta$. The second derivative is the quadratic form
\[
\Xi\ni\eta\mapsto X''_\xi[\eta]
=-2\int_0^\infty\e^{t D^\ast_\xi}
(X'_\xi[\eta]-\teta) B (X'_\xi[\eta]-\teta) \e^{t D_\xi}\d t,
\]
and concavity follows from the obvious fact that $X''_\xi[\eta]\le0$.

To prove the inequalities let us rewrite the Riccati
equation~\eqref{riccatiequation} as a Lyapunov equation
\[
X_\xi A_\xi+A^\ast_\xi X_\xi=X_\xi B X_\xi - C_\xi,
\]
and recall that, as established in the proof of Proposition~\ref{propgxi}~(5),
$\mp A_\xi$ is stable for $\pm(\xi-(2\vartheta)^{-1})>0$. Thus, we have
\[
\mp X_\xi=-\int_0^\infty\e^{\mp tA^\ast_\xi} (X_\xi BX_\xi-C_\xi)\e^{\mp t A_\xi}\d t
\le\int_0^\infty\e^{\mp tA^\ast_\xi} C_\xi\e^{\mp t A_\xi}\d t,
\]
and since $C_\xi\le0$ for $|\xi-(2\vartheta)^{-1}|\ge(2\vartheta)^{-1}$, we conclude that
$X_\xi\le0$ for $\xi<0$ and $X_\xi\ge0$ for $\xi>\vartheta^{-1}$. The controllability of
$(\mp A_\xi,Q)$ yields that these inequalities for $X_\xi$ are strict.
Writing~\eqref{riccatiequation} as
\[
X_{\xi}D_{\xi} + D^*_{\xi}X_{\xi}= - X_{\xi} B X_{\xi} - C_{\xi},
\]
the stability of $D_\xi$ gives
\[
X_\xi=\int_0^\infty\e^{tD^\ast_\xi} (X_\xi BX_\xi+C_\xi)\e^{t D_\xi}\d t\ge
\int_0^\infty\e^{tD^\ast_\xi} C_\xi\e^{t D_\xi}\d t,
\]
and since $C_\xi\ge0$ for $\xi\in\cD_0$, we can conclude that $X_\xi\ge0$
for such $\xi$. The controllability of $(D_\xi,Q)$ again yields the strict inequality.
The concavity of the map $\xi\mapsto X_\xi$ implies that the subset of all $\xi\in\cD$
such that $X_\xi>0$ is convex, so that it contains the convex hull of 
$\cD_0\cup\{\xi\in\cD\,|\,\xi>\vartheta^{-1}\}$.

From
\[
X_0=\lim_{0>\xi\to0} X_\xi\leq 0,\qquad
X_0=\lim_{\cD_0\ni\xi\to0}X_\xi\geq 0,
\]
we deduce $X_0=0$. To prove the last assertion, starting from Eq.~\eqref{riccatiequation} and
invoking Relations~\eqref{paramABCXi} one shows that
$\widehat{M}=\theta X_{\vartheta^{-1}}^{-1}\theta $ satisfies the Lyapunov equation
$A\widehat{M}+\widehat{M}A^\ast+B=0$. Since $A$ is stable, this
equation has a unique solution given by~\eqref{Mdef}, hence $\widehat{M}=M$.

\bigskip
\textbf{(4)} A simple calculation yields
\[
\mathcal{R}_{\xi+\eta}(X+\teta)=\mathcal{R}_\xi(X)+\Sigma_\teta
\]
for any $X\in L(\Gamma)$ and $\xi,\eta\in\Xi$. Thus,
$\mathcal{R}_{\xi+\eta}(X_\xi+\teta)=\Sigma_\teta$ and
since $A_{\xi+\eta}-B(X_\xi+\teta)=D_\xi$, we conclude that whenever
$\Sigma_\teta=0$ one has $X_{\xi+\eta}=X_\xi+\teta$.

\bigskip
\textbf{(5)--(7)} The proof follows line by line the one of the corresponding Parts
of~\cite[Proposition~5.5]{JPS}.

\bigskip
\textbf{(8)} Upon differentiating Eq.~\eqref{gxiForm} one gets
\[
\eta\cdot\nabla g(\xi)=-\frac12\tr\left(Q^\ast(X'_\xi[\eta]-\teta)Q\right).
\]
Further, using~\eqref{Eq:Xprime} leads to
\[
\eta\cdot\nabla g(\xi)=\frac12\int_0^\infty\tr\left(\Sigma_\teta
\e^{tD_\xi}B\e^{tD_\xi^\ast}\right)\d t,
\]
and the result now follows from Part~(5).
\hfill$\square$

\subsubsection{Proof of Proposition \ref{proplimgxi}}

By Proposition~\ref{propRiccati}, for $\xi\in\overline{\cD}$, we have
$A=D_\xi+QQ^\ast(X_\xi-\txi)$ with $\txi\rhd\xi$, and we can
rewrite the equation of the motion~\eqref{eqmotionx} as
\begin{equation}
\d x(t)= D_\xi x(t)\d t + Q\d w_\xi(t),
\label{eqmotionDxi}
\end{equation}
where
\[
w_\xi(t)=w(t)-\int_{0}^{t}Q^\ast (\txi - X_\xi)x(s)\d s.
\]
Let $Z_\xi(t)$ be the stochastic exponential of the local martingale
\[
\eta_\xi (t)=\int_0^t Q^\ast(\txi-X_\xi)x(s)\cdot\d w(s).
\]
Combining Eq.~\eqref{riccatiequation} with the relations
$\txi QQ^\ast=QQ^\ast\txi=Q\xi Q^\ast$
and $\txi QQ^\ast\txi=Q\xi^2Q^\ast$, we derive
\[
\frac12 |Q^\ast(\txi-X_\xi)x|^2=-\sigma_\txi(x)-(\txi-X_\xi)x\cdot Ax,
\]
and we can write the quadratic variation of $\eta_\xi$ as
\[
\frac12[\eta_\xi](t)=-\int_0^t\sigma_\txi(x(s))\d s
-\int_0^t(\txi-X_\xi)x(s)\cdot A x(s) \d s.
\]
Itô calculus and Proposition~\ref{propxitild}~(4) give
\[
\log Z_\xi(t)
=\eta_\xi(t)-\frac12 [\eta_\xi](t)=\xi\cdot W(t)-\chi_\xi(x(t))+\chi_\xi(x(0))-t\lambda_\xi,
\]
with
\[
\chi_\xi(x)=\frac12 x\cdot X_\xi x.
\]
and, taking~\eqref{gxiForm} into account,
\[
\lambda_\xi=-\frac12\tr(Q^\ast(X_\xi-\txi)Q)=g(\xi).
\]

The proof of the following Lemma is identical to the one of~\cite[Lemma~5.7]{JPS}, and we
omit it.

\begin{lemm}
For $\xi\in\overline{\cD}$, the process
\begin{equation}
Z_\xi(t)=\e^{-tg(\xi)+\langle\xi,\Phi(t)\rangle -\chi_\xi(x(t))+\chi_\xi(x(0))}
\end{equation}
is a $\PP_x$-martingale for all $x\in\Gamma$.
\end{lemm}

Applying Girsanov theorem, we conclude that $\{w_\xi(t)\}_{t\in [0,\tau]}$ is a
standard Wiener process under the law
$\mathbb{Q}^\tau_{\xi ,\mu}[\,\cdot\,]=\EE_\mu[Z_\xi(\tau)\,\cdot\,]$.
It follows that the finite-time cumulant generating function can be written as
\[
g_t(\xi)=tg(\xi)+\log \EE_\mu\left[ Z_\xi(t)\e^{\chi_\xi(x(t))-\chi_\xi(x(0))}\right]
=tg(\xi)+\log\mathbb{Q}^{t}_{\xi,\mu}\left[\e^{\chi_\xi(x(t))-\chi_\xi(x(0))}\right]
\]
for $\xi\in\overline{\cD}$, i.e.,
\beq
g_t(\xi)=tg(\xi)+\log d_t(\xi),\qquad
d_t(\xi)=\langle\eta^1_\xi ,Q_\xi^t\eta^2 _\xi\rangle,
\label{PreExp}
\eeq
where
\[
\eta^1_\xi(x)=\det(2\pi M)^{-\frac12}\e^{-\chi_{\xi}(x)-\frac12|M^{-\frac12}x|^2} ,
\qquad
\eta^2 _\xi(x)=\e^{\chi_{\xi}(x)},
\]
and $Q_\xi^t$ is the Markov semigroup associated with the SDE~\eqref{eqmotionDxi}.
From the explicit solution
\[
x(t)=\e^{tD_\xi}x(0)+\int_0^t\e^{(t-s)D_\xi}Q\d w_\xi(s)
\]
to this SDE we easily obtain the representation
\[
(Q^t_\xi f)(x)=\det(2\pi M_{\xi,t})^{-\frac12}
\int\e^{-\frac12 |M_{\xi,t}^{-\frac12}(y-\e^{tD_\xi}x)|^2}f(y)\d y.
\]
Setting
\[
N_{\xi,t}=\begin{bmatrix}
X_\xi+\theta X_{\vartheta^{-1}}\theta+\e^{tD_\xi^\ast}M_{\xi,t}^{-1}\e^{tD_\xi}
                  & -\e^{tD_\xi^\ast}M^{-1}_{\xi , t} \\
                  -M^{-1}_{\xi , t}\e^{tD_\xi}
                  &M_{\xi,t}^{-1}-X_\xi
\end{bmatrix},
\]
and $\cD_t=\{\xi\in\cD\,|\,N_{\xi,t}>0\}$, an elementary calculation then leads to
\begin{eqnarray*}
d_t(\xi)&=&\det(2\pi M_{\xi,t})^{-\frac12}\det(2\pi M)^{-\frac12}
\int\e^{-\frac12 z\cdot N_{\xi,t}z}\d z\\
&=&
\begin{cases}
\det( M_{\xi,t}^{-1})^{\frac12}\det( M^{-1})^{\frac12}\
\det(N_{\xi,t})^{-\frac12}&\text{for }\xi\in\cD_t;\\[4pt]
+\infty&\text{otherwise}.
\end{cases}
\end{eqnarray*}
Schur's complement formula and Proposition~\ref{propRiccati}~(5--6) lead to the factorization
\[
\det(N_{\xi,t})=d_t^-(\xi)d_t^+(\xi)
\]
where
\[
d_t^-(\xi)=\det(X_\xi+\theta X_{\vartheta^{-1}}\theta-\widetilde{\Delta}_{\xi,t}),
\qquad
d_t^+(\xi)=\det(X_{\vartheta^{-1}-\xi}+\theta\Delta_{\xi,t}\theta),
\]
$\Delta_{\xi,t}$, as defined in Proposition~\ref{propRiccati}~(6), and
$\widetilde{\Delta}_{\xi,t}=\e^{tD_\xi^\ast}(X_\xi+X_\xi
(\theta X_{\vartheta^{-1}-\xi}\theta+\Delta_{\xi,t})^{-1}X_\xi)\e^{tD\xi}$ are strictly positive for
$t>0$ and vanish exponentially as $t\to\infty$.  Thus, for $\xi\in\cD$,
\[
d^-(\xi)=\lim_{t\to\infty}d^-_t(\xi)=\det(X_\xi+\theta X_{\vartheta^{-1}}\theta),\qquad
d^+(\xi)=\lim_{t\to\infty}d^+_t(\xi)=\det(X_{\vartheta^{-1}-\xi}),
\]
and setting
\begin{align*}
\cD_t^-&=\{\xi\in\cD\,|\,
X_\xi+\theta X_{\vartheta^{-1}}\theta>\widetilde{\Delta}_{\xi,t}\},&
\cD^-&=\{\xi\in\cD\,|\,X_\xi+\theta X_{\vartheta^{-1}}\theta>0\},\\
\cD_t^+&=\{\xi\in\cD\,|\,\theta X_{\vartheta^{-1}-\xi}\theta>-\Delta_{\xi,t}\},&
\cD^+&=\{\xi\in\cD\,|\,X_{\vartheta^{-1}-\xi}>0\},
\end{align*}
one has
\[
\cD_\infty=\cD^-\cap\cD^+\subset\bigcup_{t>0}\bigcap_{s\geq t}(\cD_s^-\cap\cD_s^+).
\]
It follows that, for all $\xi\in\cD_\infty$, the limit
\[
\lim_{t\to\infty}d_t(\xi)
=\frac{\det(Y_\xi)^{1/2}\det(X_{\vartheta^{-1}})^{1/2}}
{d^-(\xi)^{1/2}d^+(\xi)^{1/2}}
\]
is finite and positive, which yields the first part of~\eqref{limgxig}. To deal with the second part,
we note that whenever
$\xi\in\Xi\setminus\overline{\cD_\infty}$, then either
$\xi\in\Xi\setminus\overline{\cD}$ or
$\xi\in\overline{\cD}\setminus\overline{\cD_\infty}$. In the latter case,
either  $X_\xi+\theta X_{\vartheta^{-1}}\theta$ or $X_{\vartheta^{-1}-\xi}$ has a negative
eigenvalue and the matrix $N_{\xi,t}$ loses its positiveness as $t\to\infty$. It follows that
$d_t(\xi)=+\infty$ and hence $g_t(\xi)=+\infty$
for large enough $t$. For $\xi\in\Xi\setminus\overline{\cD}$,
applying~\cite[Lemma~5.8]{JPS} to the functions $f_t(\alpha)=g_t(\alpha\xi)$ yields the
desired result.

Finally, we note that the continuity and concavity of the map $\cD\ni\xi\mapsto
X_\xi$ imply that $\cD_\infty$ is an open convex subset of $\cD$. The
positiveness of $N_\xi$ for $\xi\in\overline{\cD_0}\setminus\{\vartheta^{-1}\}$
is a consequence of its continuity and proves the last statement, concluding the
proof of Proposition~\ref{proplimgxi}.

\subsection{Proof of Theorem~\ref{theolocalLDP}}

Since $X_0=0$ and $X_{\vartheta^{-1}}=\theta M^{-1}\theta>0$, it follows from
the continuity of the map $\overline{\cD}\ni\xi\mapsto X_\xi$ that the open set
$\cD_\infty$ contains $0$. By Proposition~\ref{proplimgxi}, $\cD_\infty$ is the
interior of the essential domain of the limiting cumulant generating
function~\eqref{limgxig}. The Gärtner-Ellis theorem thus implies that the LDP
upper bound in~\eqref{LDF} hold for any Borel $F\subset\cL^\perp$, with the rate
function
\[
I(\varphi)=\sup_{\xi\in\cD_\infty}\left(\xi\cdot\varphi-g(\xi)\right)
=\sup_{\xi\in\cS_\infty}\left(\xi\cdot\varphi-g(\xi)\right).
\]
Moreover, the corresponding lower bound holds for any subset $F$ of the set
$\cF$ of exposed points of this function. Let us set $\cE=\nabla g(\cS_\infty)$.
We have to show that $\cF=\cE$.

By Proposition~\ref{propgxi}~(3+4+6) , for all $\varphi\in\cL^\perp$ one has
\[
0\le J(\varphi)=\sup_{\xi\in\cS}\left(\xi\cdot\varphi-g(\xi)\right)<\infty.
\]
It follows from Proposition~\ref{propgxi}~(4+7) and~\cite[Theorem~26.5]{Ro} that, as a
function on $\cL^\perp$, $J$ is the Legendre conjugate of the restriction of $g$ to $\cS$. In 
particular, it is strictly convex and differentiable on $\cL^\perp$. Moreover, 
$\nabla g:\cS\to\cL^\perp$ is a homeomorphism whose inverse is 
$\nabla J:\cL^\perp\to\cS$. Since $\cD_\infty$ is open and convex, 
so is $\cS_\infty$, and its image $\nabla g(\cS_\infty)=\nabla g(\cD_\infty)=\cE$ is open 
and connected. We note that
\[
I(\varphi)=\sup_{\xi\in\cS_\infty}\left(\xi\cdot\varphi-g(\xi)\right)\le J(\varphi)
\]
for $\varphi\in\cL^\perp$. For $\varphi=\nabla g(\xi)\in\cE$ one has
\[
J(\varphi)=\xi\cdot\nabla g(\xi)-g(\xi)=I(\varphi),
\]
i.e., $I$ and $J$ coincide on $\cE$. In particular, $I$ is strictly convex on any convex subset
of $\cE$.  Suppose that $\varphi\in\cE$ is not an exposed point of $I$. Since
$\varphi=\nabla g(\xi)$ with $\xi\in\cS_\infty$, there exists $\psi\in\cL^\perp$
such that $\psi\not=\varphi$ and $I(\psi)=I(\varphi)+\xi\cdot(\psi-\varphi)$. Invoking convexity,
one shows that $I(\psi_\lambda)=I(\varphi)+\xi\cdot(\psi_\lambda-\varphi)$ with
$\psi_\lambda=\lambda\psi+(1-\lambda)\varphi$ and $\lambda\in[0,1]$, which contradicts
the strict convexity of $I$ in a convex neighborhood of $\varphi$.

Whenever both $\pm\varphi\in\cE$, we have $\varphi=\nabla g(\xi)$ and 
$-\varphi=\nabla g(\vartheta^{-1}-\xi)$ with $\xi\in\cS_\infty$ and $\vartheta^{-1}-\xi\in\cS_\infty$ , 
and thus, 
\begin{align*}
I(-\varphi)=I(\nabla g(\vartheta^{-1}-\xi))
&=(\vartheta^{-1}-\xi)\cdot\nabla g(\vartheta^{-1}-\xi)-g(\vartheta^{-1}-\xi)\\
&=-(\vartheta^{-1}-\xi)\cdot\nabla g(\xi)-g(\xi)\\
&=I(\nabla g(\xi))-\vartheta^{-1}\cdot\nabla g(\xi)\\
&=I(\varphi)-\vartheta^{-1}\cdot\varphi.
\end{align*}
Finally, we note that $\cF_0=\{\nabla g(\xi)\,|\,\xi\in\cD_\infty\text{ and }\vartheta^{-1}-\xi\in\cD_\infty\}$. Proposition~\ref{propRiccati}~(3) implies that $X_\xi>0$ 
and $X_{\vartheta^{-1}-\xi}>0$ for $\xi\in\cD_0$, and hence that 
$\nabla g(\cD_0)\subset\cF_0$.

\subsection{Proof of Theorem~\ref{theoglobalLDP}}

By Theorem~\ref{theolocalLDP} it suffices to show that, under Condition~{\bf (R)}, 
one has $\nabla g(\cS_\infty)=\cL^\perp$. By
Proposition~\ref{propRiccati}~(4), for any $\xi\in\cS$ and any $\eta\in\cL$ one
has
\[
X_{\vartheta^{-1}-\xi-\eta}=X_{\vartheta^{-1}-\xi}-\widetilde{\eta}
\]
so that
\[
\cD^+=\{\xi\oplus\eta\,|\,\xi\in\cS,\eta\in\cL,\widetilde{\eta}<X_{\vartheta^{-1}-\xi}\}.
\]
Similarly, from
\[
X_{\xi+\eta}+\theta X_{\vartheta^{-1}}\theta
=X_\xi+\theta X_{\vartheta^{-1}}\theta+\widetilde{\eta},
\]
we deduce that
\[
\cD^-=\{\xi\oplus\eta\,|\,\xi\in\cS,\eta\in\cL,
\widetilde{\eta}>-X_\xi-\theta X_{\vartheta^{-1}}\theta\}.
\]
It follows that
\[
\cD_\infty=\{\xi\oplus\eta\,|\,
\xi\in\cS,\eta\in\cL,
-X_\xi-\theta X_{\vartheta^{-1}}\theta<\widetilde{\eta}<X_{\vartheta^{-1}-\xi}\},
\]
and hence
\[
\cS_\infty=\{\xi\in\cS\,|\,\text{there exists }\eta\in\cL\text{ such that }
-X_\xi-\theta X_{\vartheta^{-1}}\theta<\widetilde{\eta}<X_{\vartheta^{-1}-\xi}\}.
\]
Thus, under Condition~{\bf (R)}, 
$\nabla g(\cS_\infty)=\nabla g(\cS)=\nabla g(\cD)=\cL^\perp$.


\end{document}